\documentclass[journal]{IEEEtran}
\usepackage{caption2}
\usepackage{algorithm}
\usepackage{algorithmicx}
\usepackage{algpseudocode}
\usepackage{graphicx}
\usepackage[tight,footnotesize]{subfigure}
\usepackage{subfigure}
\usepackage{multicol}
\usepackage{multirow}
\usepackage{bbding}
\ifCLASSOPTIONcompsoc
  \usepackage[nocompress]{cite}
\else
  \usepackage{cite}
\fi
\begin{document}
\title{Asymmetric Correlation Quantization Hashing for Cross-modal Retrieval}

\author{Lu Wang,
        Jie Yang

\IEEEcompsocitemizethanks{\IEEEcompsocthanksitem Lu Wang was with the Institute of Image Processing and Pattern Recognition,
Shanghai Jiao Tong University, Shanghai, China,
201100.\protect\\
E-mail: luwang16@sjtu.edu.cn
\IEEEcompsocthanksitem J. Yang is  with Shanghai Jiao Tong University. E-mail: jieyang@sjtu.edu.cn}% <-this % stops an unwanted space
\thanks{Manuscript received January 13, 2020}}

% The paper headers
\markboth{IEEE TRANSACTIONS ON MULTIMEDIA}%
{Lu \MakeLowercase{\textit{et al.}}: Asymmetric Correlation Quantization Hashing for Cross-modal Retrieval}

\IEEEtitleabstractindextext{%
\begin{abstract}
Due to the superiority in similarity computation and database storage for large-scale multiple modalities data, cross-modal hashing methods have attracted extensive attention in similarity retrieval across the heterogeneous modalities. However, there are still some limitations to be further taken into account: (1) most current CMH methods transform real-valued data points into discrete compact binary codes under the binary constraints, limiting the capability of representation for original data on account of abundant loss of information and producing suboptimal hash codes; (2) the discrete binary constraint learning model is hard to solve, where the retrieval performance may greatly reduce by relaxing the binary constraints for large quantization error; (3) handling the learning problem of CMH in a symmetric framework, leading to difficult and complex optimization objective. To address above challenges, in this paper, a novel Asymmetric Correlation Quantization Hashing (ACQH) method is proposed. Specifically, ACQH learns the projection matrixes of heterogeneous modalities data points for transforming query into a low-dimensional real-valued vector in latent semantic space and constructs the stacked compositional quantization embedding in a coarse-to-fine manner for indicating database points by a series of learnt real-valued codeword in the codebook with the help of pointwise label information regression simultaneously. Besides, the unified hash codes across modalities can be directly obtained by the discrete iterative optimization framework devised in the paper. Comprehensive experiments on diverse three benchmark datasets have shown the effectiveness and rationality of ACQH.
\end{abstract}

}

\maketitle
\IEEEdisplaynontitleabstractindextext
\IEEEpeerreviewmaketitle

\section{Introduction}
\IEEEPARstart{I}{n} recent years, the ubiquity of the large amounts of multiple modalities data with high dimensional feature representation on social media and internet search engines has lead to many active fundamental topics in research in the community of machine learning \cite{proceeding1}, \cite{proceeding2}, \cite{jour14}, \cite{jour15} and computer vision \cite{jour1}, \cite{jour2}, where multimedia data always appear in three forms of image, text, and video. When performing multimedia data analysis tasks in real-word application, there are some needs of multimedia information retrieval, storage and feature compression that can work in large-scale data environment. To achieve these tasks, hashing based methods \cite{proceeding7}, \cite{proceeding10}, \cite{proceeding27}, \cite{jour13}, \cite{jour16}  has demonstrated its superiority when storing and searching large-scale data points, which transforms real-valued data points into compact binary codes under preserving the semantic similarity information and the semantic correlation structures of the original data points.

From the viewpoint of transforming strategy when encoding database points and queries, there are symmetric and asymmetric two ways in existing cross-modal hashing methods. The symmetric ones generate binary codes from the learnt unified hash functions; however, superior performance can be obtained with shorter codes in the asymmetric hashing scheme, which has been theoretically illustrated by some works \cite{proceeding3}, \cite{jour3}, \cite{jour4}. Compared with symmetric hashing methods, the asymmetric ones always use two different hash functions to generate the hash binary codes for query and database data. For example, two-step hashing \cite{proceeding4}, \cite{proceeding5}, \cite{proceeding6} decomposes the hashing learning process into two steps: a compact binary code production step and a hash functions constructing step based on the learnt binary codes, which is a classic asymmetric hashing methods when adopting different transforming strategies for database and query data. There are many advantages about two-step hashing such as superiority in performance, flexibility in encoding scheme and simplicity in computation and application.

There have been many hashing based methods developed in these years, which roughly are classified into three categories: single-modal hashing \cite{proceeding7}, \cite{proceeding8}, \cite{proceeding9}, multi-modal hashing \cite{proceeding10}, \cite{proceeding11}, \cite{jour4} and cross-modal hashing (CMH) \cite{proceeding27}, \cite{jour12}, \cite{proceeding29}, \cite{proceeding28}, \cite{proceeding22}, \cite{proceeding21}. In general, single-modal hashing only can deal with data points in one modality, which may show great limitation for multimedia in real-world scenarios. In fact, retrieving the semantic related items in some one modality database by a query in a different heterogenous modality is a desirable problem, which could be called cross-modal retrieval task. Thus, multi-modal hashing and cross-modal hashing have attracted more and more attentions, which both can achieve cross-modal similarity retrieval task. However, multi-modal hashing always needs paired-query items when searching information in database, which restricts the scope of performing cross-modal similarity retrieval task. As a result, we focus on cross-modal hashing methods for cross-modal similarity search. By whether the label information is utilized, these methods can be categorized into two subclass: unsupervised \cite{proceeding14}, \cite{proceeding15}, \cite{proceeding16}, \cite{proceeding17}, \cite{jour5} and supervised \cite{proceeding18}, \cite{proceeding19}, \cite{proceeding20}, \cite{proceeding21}, \cite{proceeding22} methods. The details are shown in the Section 2.

Although promising performance is demonstrated, it remains some limitations that need to be further taken into account. Firstly, the binary limitation is universal in current CMH methods \cite{proceeding14}, \cite{proceeding15}, \cite{proceeding16}. In other words, under the binary constraints, most existing CMH methods transform real-valued data points into discrete compact binary codes, which may limit the capability of representation strength for original data due to abundant loss of information and generate the suboptimal compact hash codes. Secondly, solving the discrete binary constraint learning model is hard. For fast solving the optimization objective, it is common to relax the discrete binary constraints and quantize the continuous result into hash codes, significantly degrading the retrieval performance with large quantization error \cite{proceeding14}, \cite{proceeding15},\cite{proceeding18}. Thirdly, learning the hash functions and binary hash codes in a symmetric form always lead to complex and difficult optimization problem, which is not applicable to training by large-scale data \cite{proceeding19}, \cite{proceeding20}, \cite{proceeding21}.

\begin {figure*}[t]
\centering
\includegraphics[width=1.0\textwidth]{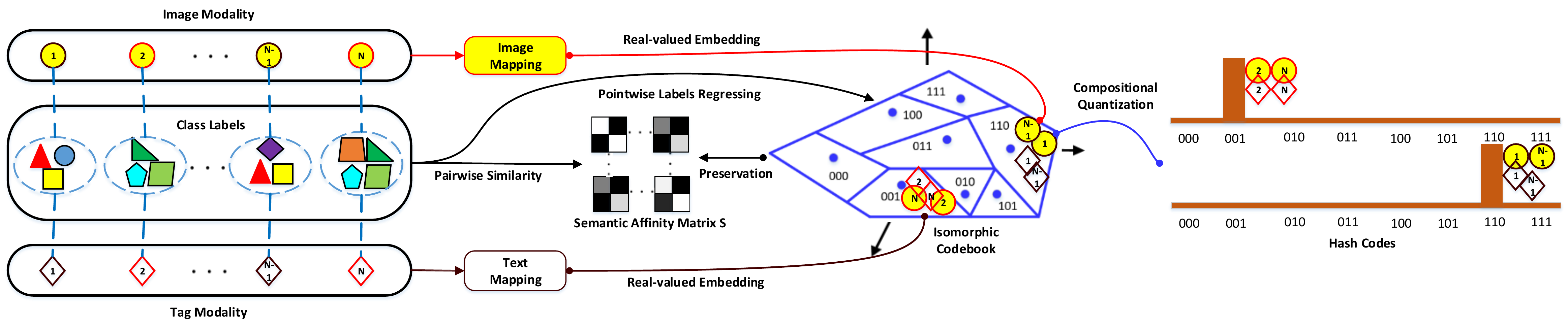}
\caption{The flowchart of ACQH.}
\label{figure100}
\end {figure*}

To address above problem, in this paper, we propose Asymmetric Correlation Quantization Hashing (ACQH), a novel hash approach achieving effective and efficient cross-modal similarity retrieval. On an intuitive level, the idea of ACQH could enable a better preservation of  the various level similarity between data points. The critical insight is that database is represented by the combination of continuous real-valued codewords in the common low-dimensional latent semantic space, i.e. the compositional quantization embedding and query is also real-valued vector in the same latent semantic space, so that much active information of the query and database points can been keep on account of breaking the binary limitation of data representation, facilitating the performance of semantic similarity search. The flowcharts of ACQH are shown in Fig. \ref{figure100}. Technically, ACQH learns the projection matrixes of heterogeneous modalities data points for transforming query into a low-dimensional real-valued vector in latent semantic space and constructs the stacked compositional quantization embedding in a coarse-to-fine manner for indicating database points by a series of learnt real-valued codeword in the codebook with the help of pointwise label information regression simultaneously.

By minimizing the difference between the ground-truth pairwise semantic similarity \cite{proceeding12}, \cite{proceeding13}, which can been obtained by the class label, and the predictive one across the heterogeneous modalities, ACQH constructs two different non-binary embeddings (i.e. the real-valued embedding for query and the stacked compositional quantization embedding for database) in a unified asymmetric hash learning framework. Furthermore, ACQH could perform a asymmetric semantic similarity calculation when retrieving queries in the searching stage. For quick retrieval, we efficiently build a lookup table for the product of the real-valued query embedding and database dictionary, which can promise the efficiency of storage and computation. However, in the way to the stacked compositional quantization, the indicator vector is constrained by binary value condition, where a mixed integer programming problem is made. In order to solve the NP-hard optimization problem, we develop a well-designed alternative optimization framework, in which each subproblem solution can be gotten handily, the quality of solutions guaranteed.

We summarize the contributions of ACQH as below.
\begin{enumerate}[]
\item We propose an asymmetric correlation quantization hashing methodology, which constructs two different non-binary embeddings for learning to hash. In the framework, we simultaneously realize the approximation of the ground-truth pairwise semantic similarity and the pointwise label information regression for each modality , leading to excellent improvement on information retrieval as the comprehensive experiments demonstrating.
\item We propose a alternately optimization framework for solving our problem. Besides, simple optimization method is developed in each step, which could result in that large-scale data can been utilized for learning efficiently.
\item Extensive experiments on three public benchmark datasets show the Superiority of ACQH for searching across the heterogeneous modalities, storing the database points and computing the asymmetric semantic similarity.
\end{enumerate}

The remainder of this paper is structured as follows. In Section 2, we briefly overview the related works of cross-modal hashing methods. Section 3 elaborates our proposed asymmetric correlation quantization hashing method, along with an efficient discrete optimization algorithm to tackle this problem. In Section 4, we report the experimental results and extensive evaluations on popular benchmark datasets. Finally, we draw a conclusion in Section 5.
\section{Related Work}
In the recent literature, there are a lot of cross-modal hashing approaches for cross-modal retrieval because of the utility and efficiency about hash technique. According to the above, existing CMH methods are grouped into two categories with unsupervised ones and supervised ones. Unsupervised CMH technique \cite{proceeding14}, \cite{proceeding15}, \cite{proceeding16}, \cite{proceeding17}, \cite{jour5} finds the hash transformation functions by keeping the relevance of the information about training data in intra-modality and inter-modality, projecting the data points into compact hash codes across the heterogeneous modalities. Meanwhile, supervised CMH \cite{proceeding18}, \cite{proceeding19}, \cite{proceeding20}, \cite{proceeding21}, \cite{proceeding22} technique usually requires semantic label information to improve the quality of the learnt hash functions and hash binary codes. Besides, end-to-end deep learning architecture has become more and more sympathetic due to its advantage in modeling the feature of original data points. Under such circumstances, it is popular to develop deep cross-modal hashing methods for their superiority in retrieval performance \cite{proceeding23}, \cite{proceeding24}.

Currently, representative unsupervised cross-modal hashing methods have LSSH \cite{proceeding14}, CMFH \cite{proceeding15}, CCQ \cite{proceeding16}, FSH \cite{proceeding17}, CRE \cite{jour5} and so on. Latent Semantic Sparse Hashing (LSSH) constructs a common abstraction semantic space to generate unified hash codes and learn corresponding hash functions for each modality by using sparse coding in image modality and matrix factorization in text modality to find the latent semantic representation \cite{proceeding14}. Collective Matrix Factorization Hashing (CMFH) searches the latent factor across the heterogeneous modalities to learn the unified binary codes by collective matrix factorization technique \cite{proceeding15}. Composite Correlation Quantization (CCQ) jointly learns the projections that could maximize the inter-modality correlation and map data points across modalities into a sharing latent semantic space and performs the composite quantization, which converts the continuous latent representation into the hash codes \cite{proceeding16}. Fusion Similarity Hashing (FSH) creates the fusion similarity across modalities by undirected asymmetric graph and embeds the fusion similarity to obtain the compact binary codes and hash functions by the graph hashing scheme \cite{proceeding17}. Collective Reconstructive Embeddings (CRE) realizes cosine similarity-based reconstructive embedding on text modality and utilizes the Euclidean distance based BRE to model image modality for learning binary codes and hash functions \cite{jour5}.

Different from unsupervised ones, CVH \cite{proceeding18}, SCM \cite{proceeding19}, SePH \cite{proceeding20}, FDCH \cite{proceeding21}, ADCH \cite{proceeding22} are representative supervised cross-modal hashing methods. Cross View Hashing (CVH) extends the single-modal spectral hashing to multiple modalities and relaxes the minimization problem for learning the hash codes \cite{proceeding18}. Semantic Correlation Maximization (SCM) learns the hash functions by approximating the semantic affinity of label information in large-scale data \cite{proceeding19}. Semantics-Preserving Hashing (SePH) builds a probability distribution by the semantic similarity of data and minimizes the Kullback-Leibler divergence for getting the hash binary codes \cite{proceeding20}. Fast Discrete Cross-modal Hashing (FDCH) learns the hash codes and hash functions with regressing from class labels to binary codes \cite{proceeding21}. Asymmetric Discrete Cross-modal Hashing (ADCH) obtains the common latent representations across the modalities by the collective matrix factorization technique to learn the hash codes and constructs hash functions by a series of binary classifiers \cite{proceeding22}.

Besides, there are also some cross-modal hashing models using deep learning methods in recent time. Some representative works are Deep Cross-Modal Hashing (DCMH) \cite{proceeding23}, Dual deep neural networks cross-modal hashing (DDCMH) \cite{proceeding24} and so on. Due to the advancement of deep learning, These deep cross-modal models have shown outstanding performance. In this paper, we mainly concentrate on shallow supervised asymmetric CMH technique.

After analyzing the existing CMH methods, there also are some challenges to address further more, i.e. the binary limitation, discrete optimization scheme with smaller quantization error and the drawback about symmetric learning form in existing CMH methods. In the next section, to solve the above problems, we will show our approach asymmetric correlation quantization hashing (ACQH), demonstrating superior cross-modal retrieval performance in our experiments.
\section{Asymmetric Correlation Quantization Hashing}
In this section, the detailed description of ACQH will been shown. We formulate ACQH in image-text bimodal data without loss of generality, which could been easily used in more modalities case with some little extension.
\subsection{Basic Formulation}
The ACQH work is supervised hashing approach, which shows superior performance in semantic similarity retrieval across modalities. Suppose that database consists of $N$ image-text pairs, where each image is represented by $d_x$-dimensional feature vector and each text is represented by $d_y$-dimensional feature vector. Let $X=\left[x_1, x_2, \ldots , x_N \right]\in R^{d_x \times N}$ be the database matrix for image modality and  $Y=\left[y_1, y_2, \ldots , y_N \right]\in R^{d_y \times N}$ be the database matrix for text modality. The database is associated with semantic labels $L=\left[l_1, l_2, \ldots , l_N \right]\in \left\{0, 1\right\}^{C \times N}$, where $C$ is the number of categories. if $x_i$ and $y_i$ belongs to the $c$-th semantic class, the $c$-th entry in $l_{i}$ equals to $1$, i.e. $l_{ci}=1$, and equals to $0$, i.e. $l_{ci}=0$ otherwise. Traditional CMH methods always construct common Hamming binary embeddings across modalities, resulting in a hash code matrix $B=\left[b_1, b_2, \ldots , b_N \right]\in \left\{-1, 1\right\}^{K \times N}$, where each data point is denoted as a binary embedding vector $b_i \in \left\{-1, 1\right\}^{K}$ and $K$ is the length of hash code.

A learning criterion of performing cross-modal Hamming binary embeddings for the obtain of these compact binary codes is to preserve the semantic affinity in original data with the Hamming affinity. Specifically, we can minimizes the difference between the ground-truth pairwise semantic similarity and the predictive one across the heterogeneous modalities for this work, a first introduced optimization model for single-modal hashing learning in kernel-based supervised hashing (KSH) \cite{proceeding7}, that is,
\begin{eqnarray}\label{CO}
\min_{B} & &  \left \|\frac{1}{K}B^TB-S \right \|_F^2 \nonumber \\
\mathrm{s.t.} & & B \in \left\{-1, 1\right\}^{K \times N},
\end{eqnarray}
where $\left \|\cdot \right \|_F$ is the Frobenius norm and $S$ is pairwise semantic similarity matrix obtained by the class label information. In above model, the binary code inner product is utilized to compute the semantic similarity between common Hamming binary embeddings, because the Hamming distance $d_H=\left| \left\{ r | b_{ir} \ne b_{jr}, 1 \leq r \leq K \right\} \right|$ equals to $\frac{1}{2} (K-b_{i}^Tb_{j})$, $b_{i}$ and $b_{j}$ being two different binary embeddings of data in the formulation. However, this formulation is constructed by the symmetric hash code inner product, which leads to some difficulty in optimization procedure and restricts the capability of approximating the real-valued pairwise semantic similarity.

Therefore, our approach Asymmetric Correlation Quantization Hashing method (ACQH) is proposed for alleviating the binary limitation, where two different non-binary embeddings (i.e. the real-valued embedding for query and the stacked compositional quantization embedding for database) is built as the extension of traditional binary values. In the following parts, the real-valued embedding for query is firstly introduced and then, the stacked compositional quantization embedding for database is explained carefully, improving the capability of pairwise semantic similarity preservation.
\subsection{Real-valued Embedding for Query}
In our ACQH, we utilize the linear hash function to project cross-modal data $x$ and $y$ into a $K$-dimensional real-valued space ($K \leq d_x$ and $K \leq d_y$), denoted by $h_x(x)=sign(W_xx)$ and $h_y(y)=sign(W_yy)$, where $W_x \in R^{d_x \times K}$ and $W_y \in R^{d_y \times K}$ are the transformation matrix for image and text modality, respectively, and  $sign(\cdot)$ is the element-wise sign quantification function. As introduced above, the real-valued embedding information for query should be used directly for  achieving more precise approximation of pairwise semantic similarity. Hence, one of the $B$s in (\ref{CO}) should be replaced by the real-valued embeddings $W_xX$ and $W_yY$ for each modality, respectively. After that, the following CMH learning model with an asymmetric structure objective is derived:
\begin{eqnarray}\label{RQ}
\min_{W_x,W_y,B} &  \left \|(W_x^TX)^TB-S \right \|_F^2 +\left \|(W_y^TY)^TB-S \right \|_F^2 \nonumber \\
\mathrm{s.t.} & B \in \left\{-1, 1\right\}^{K \times N}.
\end{eqnarray}

In this model, the semantic similarity is measured by the inner product between the real-valued embeddings $W_x^TX$ and $W_y^TY$ for each modality, and the common Hamming binary embeddings $B$, respectively. Besides, the real-valued embeddings and common Hamming binary embeddings are jointly learnt by keeping the pairwise semantic similarity. When problem (\ref{RQ}) solved, the common Hamming binary embeddings $B$ of database points can be gotten, and we can encode the unseen image query $x$ and text query $y$ into compact binary codes by the learned hash function $h_x(x)=W_xx$ and $h_y(y)=W_yy$ for each modality, respectively.
\subsection{Compositional Quantization Embedding for Database}
In model (\ref{RQ}), the learning problem is constructed in the asymmetric framework. However, there is also a limitation in above formulation, that is Hamming binary embedding for database points, capturing little various pairwise semantic similarities of database in learning procedure. To this end, we perform the stacked compositional quantization embeddings for database points compressing. As a result, the model shows great ability to preserve the semantic similarity across modality.

In our ACQH, the stacked compositional quantization technique \cite{jour8} is adopted for modeling the compositional quantization embeddings of database points. Technically, a compositional quantization embedding vector $\hat{b}_i$ of some database point is built by the product between the real-valued compositional dictionary $C$, containing several sub-dictionary, and the indicator vector $a_i$, i.e., $\hat{b}_i=Ca_i$. Here, the dictionary $C=\left[C_1, C_2, \ldots, C_m \right] \in R^{K \times mn}$ contains $m$ number of sub-dictionaries and each sub-dictionary $C_t \in R^{K \times n}$ consists of $n (n \ll N)$ number of atoms. Besides, $a_i =\left[a_i^{(1)}, a_i^{(2)}, \ldots, a_i^{(m)} \right] \in \left\{0, 1\right\}^{mn}$ is indicator vector, which is composed of $m$ number of binary $1$-of-$n$ sub-indicator vector $a_i^{(t)}$, encoded by the corresponding sub-dictionary $C_t$. In this paper, the parameter $n$ is set to $256$ \cite{proceeding25}, \cite{proceeding26}, fitting each sub-indicator vector into a byte. Based on this scheme, $\hat{B}=\left[\hat{b}_1, \hat{b}_2, \ldots, \hat{b}_N \right] \in R^{K \times N}$ has at most $n^m$ kinds of real-valued vector approximation, which enables to adequately encode various the real-valued database points into effective binary codes. In query stage, we can efficiently build a lookup table for the product of the real-valued query embedding and database dictionary for quick retrieval, which can promise the efficiency of storage and computation, and obtain the semantic similarity between query and database by table lookup operation of applying $a_i$. Accordingly, another $B$ in (\ref{RQ}) should be replaced by $\hat{B}=CA$, where $A=\left[a_1, a_2, \ldots, a_N \right]= \left[A_1; A_2; \ldots; A_m \right] \in \left\{0, 1\right\}^{mn\times N}$, and the model objective (\ref{RQ}) is reformulated as follows:
\begin{eqnarray}\label{MRQ}
\min_{W_x,W_y,C,A} & &  \! \left \|(W_x^TX)^TCA-S \right \|_F^2 \! +\! \left \|(W_y^TY)^TCA-S \right \|_F^2 \!  \nonumber \\
\mathrm{s.t.} & & A_t \in \left\{0, 1\right\}^{n \times N}, \nonumber \\
&& 1^TA_t=1^T, t=1,\ldots,m,
\end{eqnarray}
where these two non-binary embeddings could also be jointly performed as in (\ref{RQ}).
\subsection{Pointwise Label Information Regressing for Compositional Quantization Embedding}
The model objective (\ref{MRQ}) mainly utilizes the pairwise semantic relations across modality, which can be further improved by pointwise label information demonstrated in \cite{proceeding8}. Therefore, a classification error term is added to the model objective in (\ref{MRQ}) as regularization and the final objective is construct for better performance. Generally speaking, learning the mapping from the stacked compositional quantization embeddings to the class labels is a good way of using labels. Given the class labels $L$ and the stacked compositional quantization embeddings $CA$ for image database $X$ and text database $Y$, we will find the projection $V$ to bridge them, which can be described as follows:
\begin{eqnarray}\label{LR}
L=V^TCA.
\end{eqnarray}

Under this case, the existing models often minimize the least squares $\left \|L-V^TCA \right \|_F^2$ for learning the projection $V$. However, due to the least squares solution being sensitive to noise and outliers, it could add instable factors into the stacked compositional quantization embeddings $CA$ in the learning procedure. But, as proved in \cite{jour7}, the term $\left \|CA-M^TL-e_Kt^T \right \|_F^2$ regressing the class label information to the stacked compositional quantization embeddings $CA$ with a drift $t$ could bring the stability of solution, where $M$ is the projection matrix and $e_K$ is a $K$-dimension identity vector. In addition, this term can significantly cut down the computation complexity in optimization process, represented in next subsection. Hence, the learning model can be formulated by minimizing this term, that is:
\begin{eqnarray}\label{LRR}
\min_{C,A,M,t} & &  \left \|CA-M^TL-e_Kt^T \right \|_F^2 + \mu \left \|M \right \|_F^2 \nonumber \\
\mathrm{s.t.} & & A_t \in \left\{0, 1\right\}^{n \times N}, \nonumber \\
& & 1^TA_t=1^T, t=1,\ldots,m,
\end{eqnarray}
where $\mu$ is a regularization parameter.

For the CMH goals, the final objective function of ACQH has following formulation:
\begin{eqnarray}\label{ACQ}
\min_{W_x,W_y,C,A,M,t} & & \left \|(W_x^TX)^TCA-S \right \|_F^2 +  \nonumber \\
&&\left \|(W_y^TY)^TCA-S \right \|_F^2 \nonumber \\
&& + \lambda \left \|CA-M^TL-e_Kt^T \right \|_F^2 + \mu \left \|M \right \|_F^2 \nonumber \\
\mathrm{s.t.} & &A_t \in \left\{0, 1\right\}^{n \times N}, \nonumber \\
& & 1^TA_t=1^T, t=1,\ldots,m,
\end{eqnarray}
where $\lambda$ is a penalty parameter which fixes the strength of pointwise label information regressing. Once model (\ref{ACQ}) is optimized, the stacked compositional quantization embeddings $CA$ can be utilized to compress the database items. Further more, the linear hash function can encode the unseen queries across modality.
\section{Algorithms Design and Analysis}
\subsection{Algorithms Design}
Generally, the mixed binary programming objective (\ref{ACQ}) is non-convex with variables $W_x$, $W_y$, $C$, $A$, $M$ and $t$ together. To address this issue, we develop an alternative iterative optimization framework. In the optimization procedure, only one variable is programmed with the other  variable fixed at each step. The details of the alternative optimization scheme are shown as follows:
\begin{enumerate}[]

\item $W_x$ and $W_y$ step.

By fixing $C$, $A$, $M$ and $t$, problem (\ref{ACQ}) is then facilitated as:
\begin{eqnarray}\label{ACQ1}
\min_{W_x} & &  \left \|(W_x^TX)^TCA-S \right \|_F^2,
\end{eqnarray}
which can easily find the solution by using matrix manipulations, leading to a closed-form solution for $W_x$:
\begin{eqnarray}\label{ACQ2}
W_x=(XX^T)^{-1}XS(CA)^T(CA(CA)^T)^{-1}.
\end{eqnarray}

By the same way, we can get the another closed-form solution for $W_y$:
\begin{eqnarray}\label{ACQ3}
W_y=(YY^T)^{-1}YS(CA)^T(CA(CA)^T)^{-1}.
\end{eqnarray}

\item $M$ step.

By fixing $W_x$, $W_y$, $C$, $A$ and $t$, problem in Eq.(\ref{ACQ}) is reformulated as:
\begin{eqnarray}\label{ACQ4}
\min_{M} & &  \left \|CA-M^TL-e_Kt^T \right \|_F^2 + \frac{\mu}{\lambda} \left \|M \right \|_F^2.
\end{eqnarray}

By setting the derivative of Eq.(\ref{ACQ4}) with respect to $M$ to zero, $M$ can be obtained with a closed-form solution:
\begin{eqnarray}\label{ACQ5}
M=(LL^T+\frac{\mu}{\lambda}I)^{-1}L(e_Kt^T-CA)^T.
\end{eqnarray}

\item $t$ step.

We solve the drift $t$ with the other model variables fixed. Hence, the drift $t$ can be represented as:
\begin{eqnarray}\label{ACQ6}
t=\frac{(CA-M^TL)^Te_K}{K}.
\end{eqnarray}

\item $C$ step.

By fixing $W_x$, $W_y$, $A$, $M$ and $t$, taking the partial derivative of the model objective defined by (\ref{ACQ}) with respect to $C$, we can acquire the closed-form solution of $C$ by setting it to zero as:
\begin{eqnarray}\label{ACQ7}
C=&(W_x^TXX^TW_x+W_y^TYY^TW_y+\lambda I)^{-1} \nonumber \\
&(W_x^TXS+W_y^TYS+\lambda P)A^T(AA^T)^{-1},
\end{eqnarray}
where $P=M^TL+e_Kt^T$.

\item $A$ step.

By fixing $W_x$, $W_y$, $C$, $M$ and $t$, the objective problem (\ref{ACQ}) is then simplified as:
\begin{eqnarray}\label{ACQ8}
\min_{A} & &  \left \|S-C^xA \right \|_F^2 +\left \|S-C^yA \right \|_F^2 \nonumber \\
& & + \lambda \left \|P-CA\right \|_F^2  \nonumber \\
\mathrm{s.t.} & & A_t \in \left\{0, 1\right\}^{n \times N}, \nonumber \\
& & 1^TA_t=1^T, t=1,\ldots,m,
\end{eqnarray}
where $P=M^TL+e_Kt^T$, $C^x=(W_x^TX)^TC$ and $C^y=(W_y^TY)^TC$. Besides, for efficiently solving the problem (\ref{ACQ8}), we can write $C^x$ and $C^y$ as $C^x=\left[C_1^x, C_2^x, \ldots, C_m^x \right]$ and $C^y=\left[C_1^y, C_2^y, \ldots, C_m^y \right]$, respectively, where $C_i^x=(W_x^TX)^TC_i$, $C_i^y=(W_y^TY)^TC_i$ and $1 \leq i \leq m$.

Then, we sequentially learn the binary indicator matrix $A$ in a coarse-to-fine style like the stacked quantization method \cite{jour8}. Concretely, the residual matrices of image modality, text modality and classification error are denoted by $R_t^x$, $R_t^y$ and $R_t$ after the $(t-1)$-th round learning, where the formulation is described as follow:
\begin{eqnarray}\label{ACQ9}
&&R_1^x=S, R_1^y=S, R_1=P; \nonumber \\
&&R_t^x=R_{t-1}^x-C_{t-1}^xA_{t-1}, R_t^y=R_{t-1}^y-C_{t-1}^yA_{t-1}, \nonumber \\ &&R_t=R_{t-1}-C_{t-1}A_{t-1}; \nonumber \\
&&t=1, 2, \ldots, m.
\end{eqnarray}

Then, these three residual matrices are further represented by the $t$-th sub-codebook $C_t^x$, $C_t^y$ and $C_t$ to learn the optimal binary matrix $A_t=\left[a_1^{(t)}, a_2^{(t)}, \ldots, a_N^{(t)} \right] \in \left\{0, 1\right\}^{n \times N}$:
\begin{eqnarray}\label{ACQ10}
\min_{A_t} & &  \left \|C_t^xA_t-R_t^x \right \|_F^2 +\left \|C_t^yA_t-R_t^y \right \|_F^2 \nonumber \\
& & + \lambda \left \|C_tA_t-R_t\right \|_F^2  \nonumber \\
\mathrm{s.t.} & & A_t \in \left\{0, 1\right\}^{n \times N}, 1^TA_t=1^T.
\end{eqnarray}

After $m$ times recursively learning, the unified binary indicator codes $A=\left[A_1; A_2; \ldots; A_m \right]$ for CMH search can be acquired. Note that, (\ref{ACQ10}) can be made a solution in brief by brute-force method because of every $a_i^{(t)}$ in $A_t$ being a 1-of-K indicator vector.
Specifically, $A_t$ can be solved in the way of column by column, due to separability of the sub-problems (\ref{ACQ10}) with regard to the columns of $A_t$. Hence, once some column $a$ of $A_t$ is solved, we can formulate the corresponding sub-problem as:
\begin{eqnarray}\label{ACQ11}
\min_{a} & &  a^TMa-2a^Th \nonumber \\
\mathrm{s.t.} & & a \in \left\{0, 1\right\}^{n}, \left \|a \right \|_0 =1,
\end{eqnarray}
where $\left \|\cdot \right \|_0$ is the $l_0$ norm, $M=(C_t^x)^TC_t^x+(C_t^y)^TC_t^y+\lambda (C_t)^TC_t$, and $h=(C_t^x)^Tr_t^x+(C_t^y)^Tr_t^y+(C_t)^Tr_t$. Besides, $r_t^x$, $r_t^y$ and $r_t$ are the corresponding columns to $a$ in $R_t^x$, $R_t^y$ and $R_t$, respectively. Then, we handle the discrete binary optimization problems (\ref{ACQ11}) directly, showing better performance. The final solution of $a$ can be simply acquired:
\begin{equation}\label{ACQ12}
a_i=
\begin{cases}
1, \mbox{if $i=j$}; \\
0, \mbox{otherwise};
\end{cases}
\end{equation}
where $j=\arg \min_{j=1,\ldots,n} M_{jj}-2h_j$.

\end{enumerate}

For clarity of reading, the whole alternative iterative optimization scheme of ACQH is summed up in Algorithm \ref{alg:ACQH}.

\begin{algorithm}[h]
  \caption{Asymmetric Correlation Quantization Hashing.}
  \label{alg:ACQH}
  \begin{algorithmic}[1]
    \Require
      feature matrices $X$ and $Y$, label matrix $L$, and similarity matrix $S$; code length $K$, sub-dictionary size $n$, number of sub-codebooks $m$, maximum iteration number $T$; parameters $\lambda$, $\mu$.
    \Ensure
      hash codes $C$, $A$, $W_x$ and $W_y$.
    \State Initialize $M$ randomly.
    \State Initialize $C$ randomly.
    \State Initialize binary indicator matrix $A$ randomly according $C$ by NN search.
    \State Initialize the drift $t=0$.
    \Repeat
      \State Update $W_p$, $(p \in \left\{x, y\right\})$, by Eq.(\ref{ACQ2}) and Eq.(\ref{ACQ3}).
      \State Update $M$ by Eq.(\ref{ACQ5}).
      \State Update $t$ by Eq.(\ref{ACQ6}).
      \State Update $C$ by Eq.(\ref{ACQ7}).
      \State Update binary indicator matrix $A$ column by column using Eq.(\ref{ACQ12}) .
    \Until Objective function of Eqn.(\ref{ACQ}) converges.
  \end{algorithmic}
\end{algorithm}
\subsection{Query}
After above learning process, we can get the projection matrix $W_x, W_y$ for image modality and text modality respectively, the common real-valued compositional codebook $C$, and the unified binary indicator matrix $A$. When searching items in database for the query, we first should compute the semantic similarity between query and database across modality, and then, the similar items to query can be found by this semantic similarity for cross-modal information retrieval task. Specifically, for a $d_x$-dimensional image modality query $q_x$ or a $d_y$-dimensional text modality query $q_y$, we first obtain the projection $\hat{q_x}$ or $\hat{q_y}$ in the $K$-dimensional latent semantic space for each modality respectively by the linear hash functions $\hat{q_x}=W_x^Tq_x$ or $\hat{q_y}=W_y^Tq_y$, respectively. Next, the semantic similarity across modality between the common real-valued compositional quantization embeddings $CA$ of the database items and $\hat{q_x}$ or $\hat{q_y}$ can be easily computed, i.e. $\hat{S}=(\hat{q_x})^TCA$ or $\hat{S}=(\hat{q_y})^TCA$, respectively. Once the semantic similarity across modality $\hat{S}$ acquired, the cross-modal approximate nearest neighbor (ANN) for the query can be found after sorting this semantic similarity.

To promise the efficiency of storage and computation, we efficiently build a lookup table for the product of the real-valued query embedding $\hat{q_x}$ or $\hat{q_y}$ and the common database dictionary $C$, denoted by $T_c=(\hat{q_x})^TC$ or $T_c=(\hat{q_y})^TC$, respectively. As for the calculation of $\hat{S}=T_cA$, we only need to perform the table lookup operations and addition operations by the guidance of the binary code matrix $A \in \left\{0, 1\right\}^{mn\times N}$ of database points.
\subsection{Analysis}
$\textbf{Query complexity.}$ In this part, we show the analysis of ACQH query complexity for retrieving cross-modal items in database. The time complexity for a new query is $O (mN)$, because we only compute the lookup table $T_c$, consuming $O (Kn)$ time, and calculate the semantic similarity $T_cA$ by the table lookup operations and addition operations, consuming $O (mN)$ time, in the query stage. Besides, $Kn \ll N$ leads to that the size $N$ of retrieval database points and the number of sub-codebooks $m$ mainly control the query complexity. Hence, the time complexity of query has a linear relation to the database size $N$, which has nothing to do with the binary code length in CMH retrieval tasks. This is another merit of ACQH, facilitating long binary code representation for better performance, which is different to the traditional CMH methods. Further more, the time complexity for training phase will be carefully discussed in next section.

$\textbf{Storage complexity.}$ The space complexity of ACQH is satisfactory, due to the simplicity in search stage. Specifically, we need to store the common dictionary $C \in R^{K \times mn}$ and the database binary indictor matrix $A \in \left\{0, 1\right\}^{mn\times N}$. Owing to the size of common dictionary $C$ relatively small, we usually ignore the memory cost of common dictionary $C$. As for the binary indictor $a$, we only need to store the particular index of $1$ in every $a$ for recording each database point. Specifically, each item (i.e. $a$) may cost $m \log_2 (n)$ bits memory. As a result, the memory of ACQH only depends on the number of sub-codebooks $m$ and the sub-dictionary size $n$, rather than the code length $K$. In practice, $m$ and $n$ are always set to a fixed values, leading to less memory cost than the traditional CMH methods.
\section{Experiments}
In this section, ACQH is compared with other state-of-the-art CMH methods on three multi-modal benchmark data sets in terms of retrieval performance. All the results are performed on a 64-bit windows PC with 2.20GHz i7-8750H CPU and 16.0GB RAM.
\subsection{Experimental Settings}
$\textbf{Datasets.}$ The experiments evaluating the performance of ACQH is carried out on three multi-modal benchmarks datasets: Wiki \cite{jour9}, MIRFlickr25K \cite{jour10} and NUS-WIDE \cite{jour11}. Specifically, some descriptions of statistics about these three datasets are presented as following.

Wiki \cite{jour9} comprises $2,866$ image-text pairs from Wikipedia’s articles. There are $10$ semantic categories, where each image-text pair has one label of the $10$ semantic categories. In this dataset, a $128$-dimensional bag-of-words visual vector is used to describe each image, and  a $10$-dimensional topics vector is used to describe each text. In Wiki, randomly selecting $2,173$ image-text pairs from the whole dataset is utilized as the training set, which also is used as the database in retrieval evaluation, and the remaining $693$ image-text pairs is utilized as the query dataset. In the evaluation, we define two samples are semantically related through the shared labels.

MIRFlickr25K \cite{jour10} consists of $25,000$ images with related tags, having $24$ semantic categories. Then, we only pick out the image-text pairs, whose tags appear at least $20$ times. Hence, a multi-modal dataset of $20,015$ image-tag pairs can be constructed, in which a $150$-dimensional histogram vector is used to represented the image, and $500$-dimensional latent semantic vector is used to represented text. In MIRFlickr25K, randomly taking $2000$ image-tag pairs from the whole dataset is utilized the query set, and the training dataset contains the left pairs, serving as the database in retrieval task. In the evaluation, we define two samples are semantically related through the shared labels.

NUS-WIDE \cite{jour11} includes $269,648$ images, which is grouped into  $81$ concepts. Following \cite{jour11}, only retaining the $10$ most frequent concepts, $186,577$ image-tag pairs can be left in a new dataset, where $500$-dimensional bag-of-words visual vector is used to represent image, and $1,000$-dimensional index vector is used to represent text. In NUS-WIDE, randomly picking out $2,000$ image-tag pairs from the new dataset is serving as the query set, and the database contains the remaining $184,577$ image-tag pairs, which is also utilized as the training set. In the evaluation, we define two samples are semantically related through the shared labels.

$\textbf{Compared methods.}$ Owing to our approach ACQH being supervised CMH, for a fair comparison, we compare ACQH with several state-of-the-art supervised CMH approaches: GSPH \cite{proceeding27}, DCH \cite{jour12}, DCMH \cite{proceeding29}, QCH \cite{proceeding28}, ADCH \cite{proceeding22}, FDCH \cite{proceeding21}. In the experiments, according to the corresponding papers, we set the training dataset and the parameters in above CMH methods. Besides, under random data partitions, we repeatedly carry out $5$ times all experiments and compare the averaged results.
\subsection{Experimental Details}
$\textbf{Pairwise Semantic Similarity Construction.}$ When the class label vectors given, we could construct pairwise semantic similarity matrix $S$ by $L^TL$ for better use of semantic information \cite{proceeding5}. Besides, using the factorization $L^TL$ replaces pairwise semantic similarity $S$ in the learning procedure leads to the reduction of computation complexity from $O (N^2)$ to $O (NC)$.

$\textbf{Implementation details.}$ For the baseline methods, we employ the public codes and the fixed parameters from the corresponding papers. As for ACQH, there are three model parameters: the regularization hyper-parameter $\mu$, the label regressing error hyper-parameter $\lambda$, the number of sub-dictionaries hyper-parameter numCodebook. In this paper, the regularization hyper-parameter $\mu$ is set to $0.01$,  the label regressing error hyper-parameter $\lambda$ is set to $0.0001$, and the number of sub-dictionaries hyper-parameter numCodebook is set to $4$ throughout the experiments. The parameter sensitivity will be studied in after section, checking the stability with a wide range of parameter.

$\textbf{Evaluation criteria.}$ In the evaluation, there are three widely used metrics: mean average precision (MAP), topN-precision curves and precision-recall curves, showing the information retrieval performance. These evaluation criteria can be closely found in \cite{proceeding30}. Additionally, the training cost with different training set size and the query cost with various code lengths are also been discussed. For all methods, query encoding and similarity computing are the all parts in the query cost. Further more, we evaluate our approach on two typical cross-modal tasks: image-query-text task (i.e. I$\rightarrow$T) and text-query-image task (i.e. T$\rightarrow$I).
\subsection{Experimental Results}
\renewcommand{\arraystretch}{1.0}
\begin{table}[htb]
  \centering
  \footnotesize
  %\fontsize{6.5}{8}\selectfont
  \setlength{\belowcaptionskip}{10pt}
  \caption{Comparison of MAP with Two Multi-modal Retrieval Tasks on Wiki Benchmark.}
  \label {Table.1}
  \resizebox{0.45\textwidth}{!}{
    \begin{tabular}{|c|c|c|c|c|c|c|c|}
    \hline
    \multirow{2}{*}{Tasks}& \multirow{2}{*}{Method} &
    \multicolumn{5}{c|}{Wiki}\cr\cline{3-7}
    &&8 bits&16 bits&32 bits&64 bits&128 bits\cr
    \hline
\hline
    \multirow{6}{*}{I$\rightarrow$T}
    &GSPH &0.2399& 0.2872&	0.2773&	0.2958&	0.2994
\cr\cline{2-7}
   &DCH &0.1937& 0.2138&	0.2391&	0.2466&	0.2479
\cr\cline{2-7}
    &DCMH &0.1238& 0.1176&	0.1164&	0.1266&	0.1233

\cr\cline{2-7}
    &QCH &0.2185& 0.2395&	0.2304&	0.2271&	0.1998
\cr\cline{2-7}
    &ADCH &0.1179& 0.1179&	0.1179&	0.1179&	0.1179
\cr\cline{2-7}
    &FDCH &0.1298& 0.1235&	0.1407&	0.1344&	0.1448
\cr\cline{2-7}
    &{\bf ACQH} &{\bf0.3126}& {\bf0.3230}& {\bf0.3239}& {\bf0.3181}& {\bf0.3228}
\cr
    \hline
\hline
    \multirow{6}{*}{T$\rightarrow$I}
    &GSPH &0.4445& 0.5017&	0.5166&	0.5275&	0.5242
\cr\cline{2-7}
   &DCH &0.2435& 0.2795&	0.3249&	0.3401&	0.3486
\cr\cline{2-7}
    &DCMH &0.1174& 0.1184&	0.1198&	0.1164&	0.1186
\cr\cline{2-7}
    &QCH &0.2645& 0.3191&	0.3128&	0.2905&	0.2025
\cr\cline{2-7}
    &ADCH &0.1179& 0.1179&	0.1179&	0.1179&	0.1179
\cr\cline{2-7}
    &FDCH &0.2336& 0.2901&	0.3238&	0.3305&	0.3425
\cr\cline{2-7}
    &{\bf ACQH} &{\bf0.6898}& {\bf0.7089}& {\bf0.7098}& {\bf0.7007}& {\bf0.7080}
\cr
    \hline
    \end{tabular}}
\end{table}

\renewcommand{\arraystretch}{1.0}
\begin{table}[htb]
  \centering
  \footnotesize
  %\fontsize{6.5}{8}\selectfont
  \setlength{\belowcaptionskip}{10pt}
  \caption{Comparison of MAP with Two Multi-modal Retrieval Tasks on MIRFlickr25K Benchmark.}
  \label {Table.2}
  \resizebox{0.45\textwidth}{!}{
    \begin{tabular}{|c|c|c|c|c|c|c|c|}
    \hline
    \multirow{2}{*}{Tasks}& \multirow{2}{*}{Method} &
    \multicolumn{5}{c|}{MIRFlickr25K}\cr\cline{3-7}
    &&8 bits&16 bits&32 bits&64 bits&128 bits\cr
    \hline
    \hline
    \multirow{6}{*}{I$\rightarrow$T}
    &GSPH &0.6547& 0.6761&	0.6926&	0.6976&	0.7059
\cr\cline{2-7}
   &DCH &0.6518& 0.6651&	0.6730&	0.6745&	0.6954
\cr\cline{2-7}
    &DCMH &0.5717& 0.5649&	0.5779&	0.5695&	0.5716
\cr\cline{2-7}
    &QCH &0.6411& 0.6614&	0.6537&	0.6581&	0.6672
\cr\cline{2-7}
    &ADCH &0.6504& 0.6587&	0.6722&	0.6846&	0.6458
\cr\cline{2-7}
    &FDCH &0.6411& 0.6800&	0.6739&	0.6770&	0.6816
\cr\cline{2-7}
    &{\bf ACQH} &{\bf0.8105}& {\bf0.8105}&	{\bf0.8097}&	{\bf0.8076}&	{\bf0.8087}
\cr
    \hline
\hline
    \multirow{6}{*}{T$\rightarrow$I}
    &GSPH &0.6580& 0.6967&	0.7175&	0.7320&	0.7391
\cr\cline{2-7}
   &DCH &0.6553& 0.6561&	0.6787&	0.6992&	0.7099
\cr\cline{2-7}
    &DCMH &0.5192& 0.5664&	0.5647&	0.5649&	0.5695
\cr\cline{2-7}
    &QCH &0.6204& 0.6353&	0.6472&	0.6556&	0.6665
\cr\cline{2-7}
    &ADCH &0.6407& 0.6633&	0.6804&	0.6898&	0.6460
\cr\cline{2-7}
    &FDCH &0.6174& 0.6556&	0.6680&	0.6746&	0.6833
\cr\cline{2-7}
    &{\bf ACQH} &{\bf0.9113}& {\bf0.9114}& {\bf0.9118}& {\bf0.9115}& {\bf0.9119}
\cr
    \hline
    \end{tabular}}
\end{table}

\renewcommand{\arraystretch}{1.0}
\begin{table}[htb]
  \centering
  \footnotesize
  \setlength{\belowcaptionskip}{10pt}
  \caption{Comparison of MAP with Two Multi-modal Retrieval Tasks on NUS-WIDE Benchmark.}
  \label {Table.3}
  \resizebox{0.45\textwidth}{!}{
    \begin{tabular}{|c|c|c|c|c|c|c|c|}
    \hline
    \multirow{2}{*}{Tasks}& \multirow{2}{*}{Method} &
    \multicolumn{5}{c|}{NUS-WIDE}\cr\cline{3-7}
    &&8 bits&16 bits&32 bits&64 bits&128 bits\cr
    \hline
    \hline
    \multirow{7}{*}{I$\rightarrow$T}
    &GSPH &0.4992& 0.5482&	0.5394&	0.5570&	0.5726
\cr\cline{2-7}
   &DCH &0.4520& 0.4758&	0.4933&	0.5069&	0.5250
\cr\cline{2-7}
    &DCMH &0.3488& 0.3200&	0.3308&	0.3299&	0.3321
\cr\cline{2-7}
    &QCH &0.4645& 0.4408&	0.4350&	0.4082&	0.3807
\cr\cline{2-7}
    &ADCH &0.5686& 0.5461&	0.5596&	0.3154&	0.4937
\cr\cline{2-7}
    &FDCH &0.3154& 0.3154&	0.3154&	0.3154&	0.3154
\cr\cline{2-7}
    &{\bf ACQH} &{\bf0.7948}& {\bf0.7909}&	{\bf0.7893}&	{\bf0.7973}&	{\bf0.7968}
\cr
    \hline
\hline
    \multirow{7}{*}{T$\rightarrow$I}
    &GSPH &0.4126& 0.5135&	0.4756&	0.5491&	0.5511
\cr\cline{2-7}
   &DCH &0.4763& 0.5049&	0.5324&	0.5485&	0.5717
\cr\cline{2-7}
    &DCMH &0.3253& 0.3387&	0.3218&	0.3172&	0.3173
\cr\cline{2-7}
    &QCH &0.4304& 0.4425&	0.4169&	0.3942&	0.3726
\cr\cline{2-7}
    &ADCH &0.5082& 0.4929&	0.4963&	0.3154&	0.4380
\cr\cline{2-7}
    &FDCH &0.3154& 0.3154&	0.3154&	0.3154&	0.3154
\cr\cline{2-7}
    &{\bf ACQH} &{\bf0.9069}& {\bf0.9047}&	{\bf0.9001}& {\bf0.9068}& {\bf0.9077}
\cr
    \hline
    \end{tabular}}
\end{table}

$\textbf{Retrieval performance.}$ We first show the MAP evaluation results with different code length from $8$ bits to $128$ bits of our ACQH and all six baselines for cross-modal search tasks on all three datasets (i.e. Wiki, MIRFlickr25K, and NUS-WIDE) in Table \ref{Table.1}, \ref{Table.2}, and \ref{Table.3}, respectively. First, as illustrated in above tables, our approach ACQH remarkably outperforms the all six baseline CMH methods on all three datasets across various code lengths for both two cross-modal information search tasks. Specifically, our ACQH is superior to the second best CMH method GSPH, outperforming GSPH averagely by $4.02\%$, $12.40\%$ and $25.05\%$ for image-query-text search task, and $20.05\%$, $20.29\%$ and $40.49\%$ for text-query-image search task on Wiki, MIRFlickr25K, and NUS-WIDE benchmark datasets respectively. From this observation, the greatest retrieval performance boosts of the proposed ACQH can be obtained. Second, from above tables, we can find that our ACQH can also achieve preferable performance over the all six baseline CMH methods with shorter code length , showing that the retrieval performance is not nearly change with various code length from $8$ bits to $128$ bits. This result represents that ACQH can gain satisfactory retrieval items even with short hash codes by using the real-valued embedding for query and the stacked compositional quantization embedding for database. Third, another observation can be discovered, where the average advance for text-query-image search task is larger than the average advance for image-query-text search task. This phenomenon may be that image modality always contains more noise and outliers that text modality. Finally, we can deduce a conclusion that the asymmetric framework seamlessly integrating the real-valued embedding for query and the stacked compositional quantization embedding for database leads to the superiority of ACQH.

\begin{figure} [t]
\centering
\subfigure{
  \includegraphics[width=.2\textwidth]{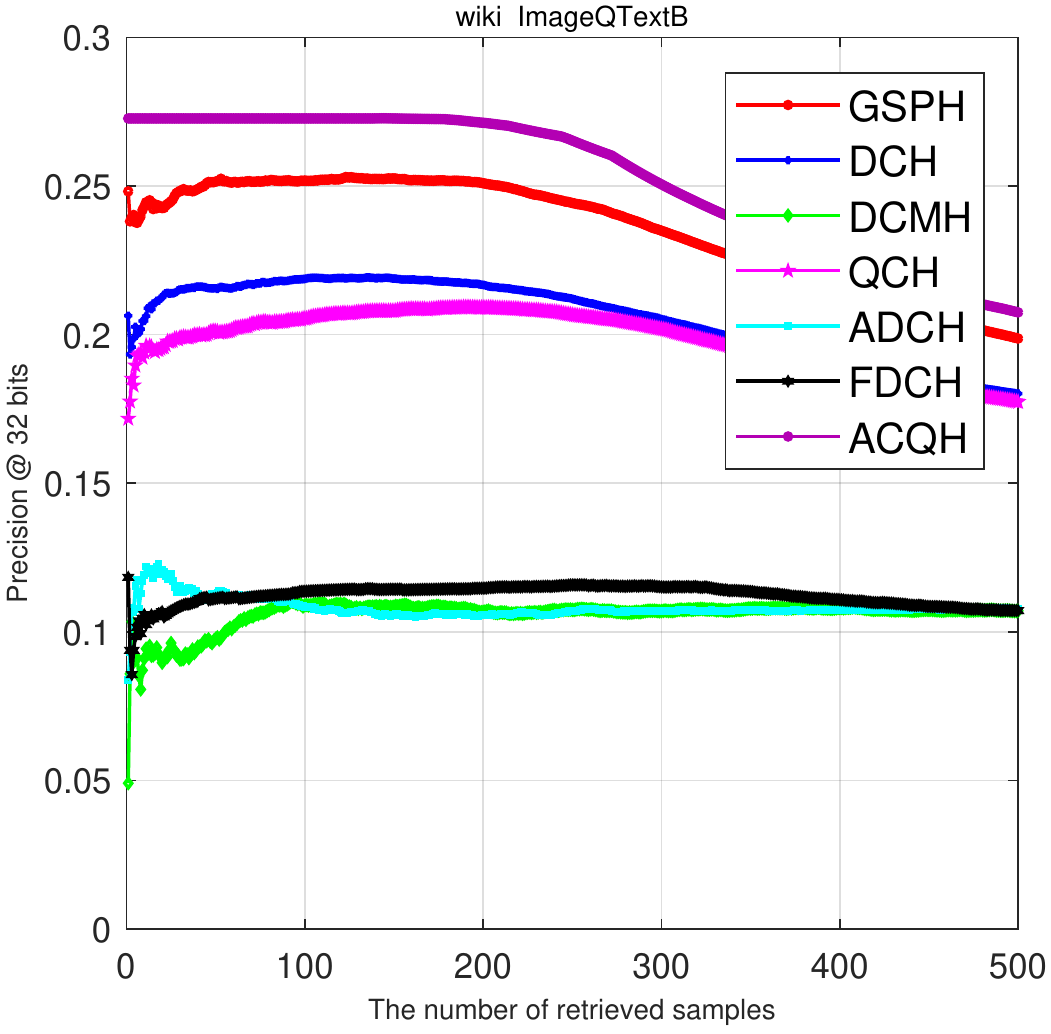}}
\hfill
\centering
\subfigure{
  \includegraphics[width=.2\textwidth]{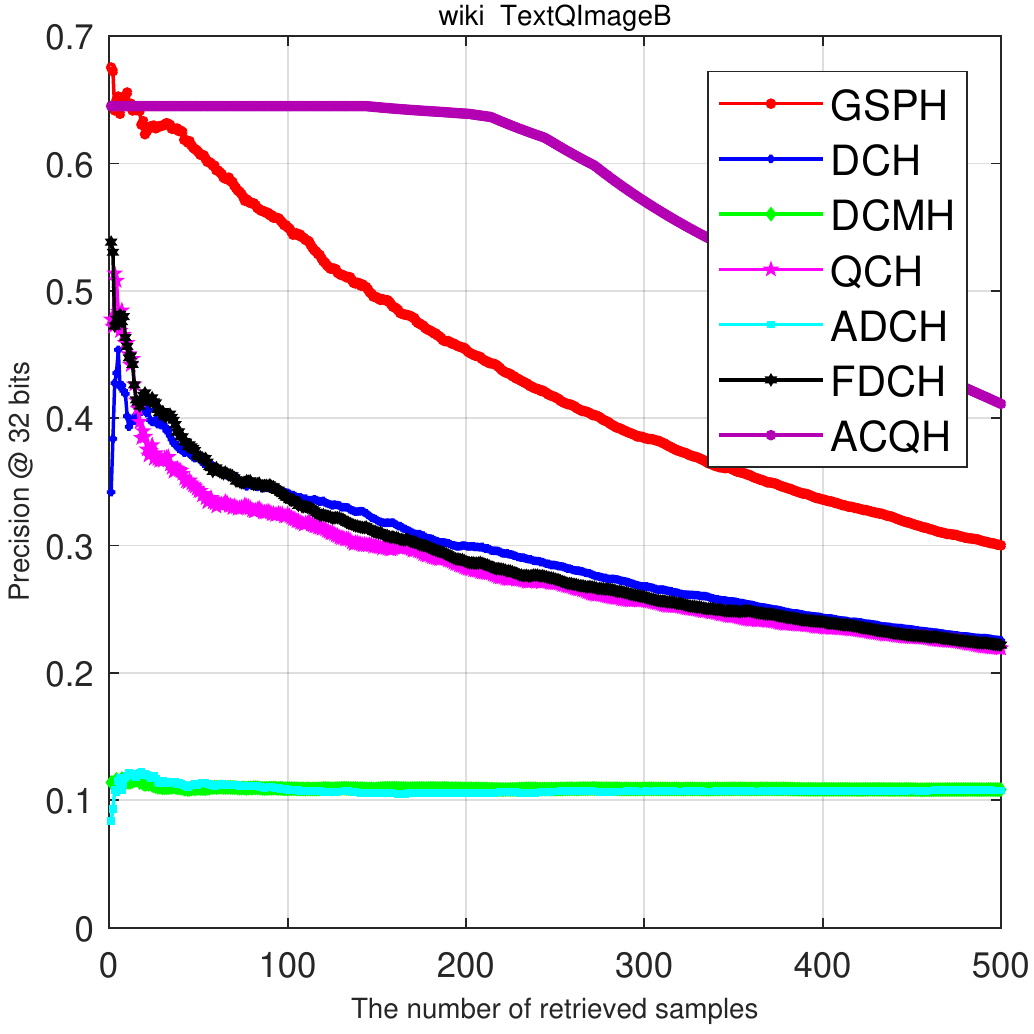}}
\captionsetup{justification=centering}
\caption{TopN-precision Curves @ $32$ bits on Wiki.}
\label{figure13}
\end{figure}

\begin{figure} [t]
\centering
\subfigure{
  \includegraphics[width=.2\textwidth]{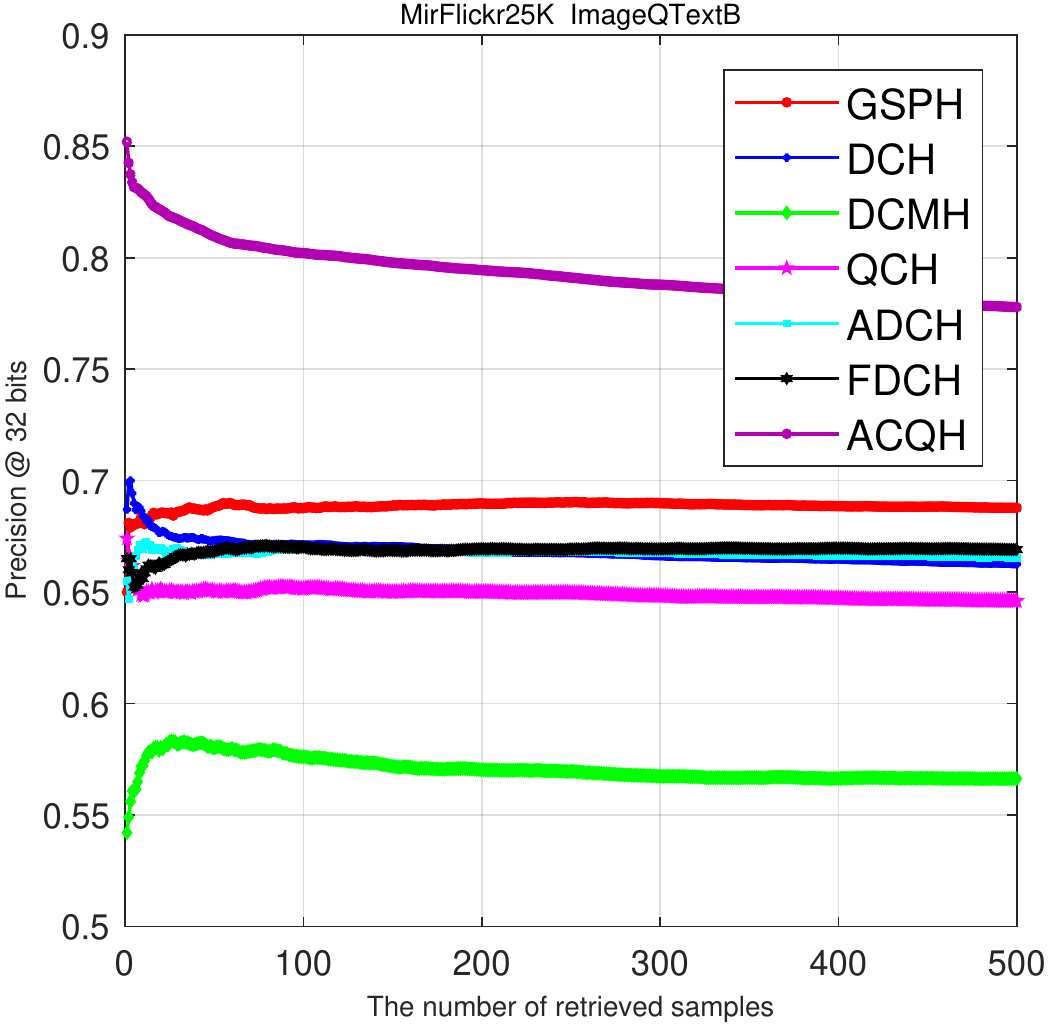}}
\hfill
\centering
\subfigure{
  \includegraphics[width=.2\textwidth]{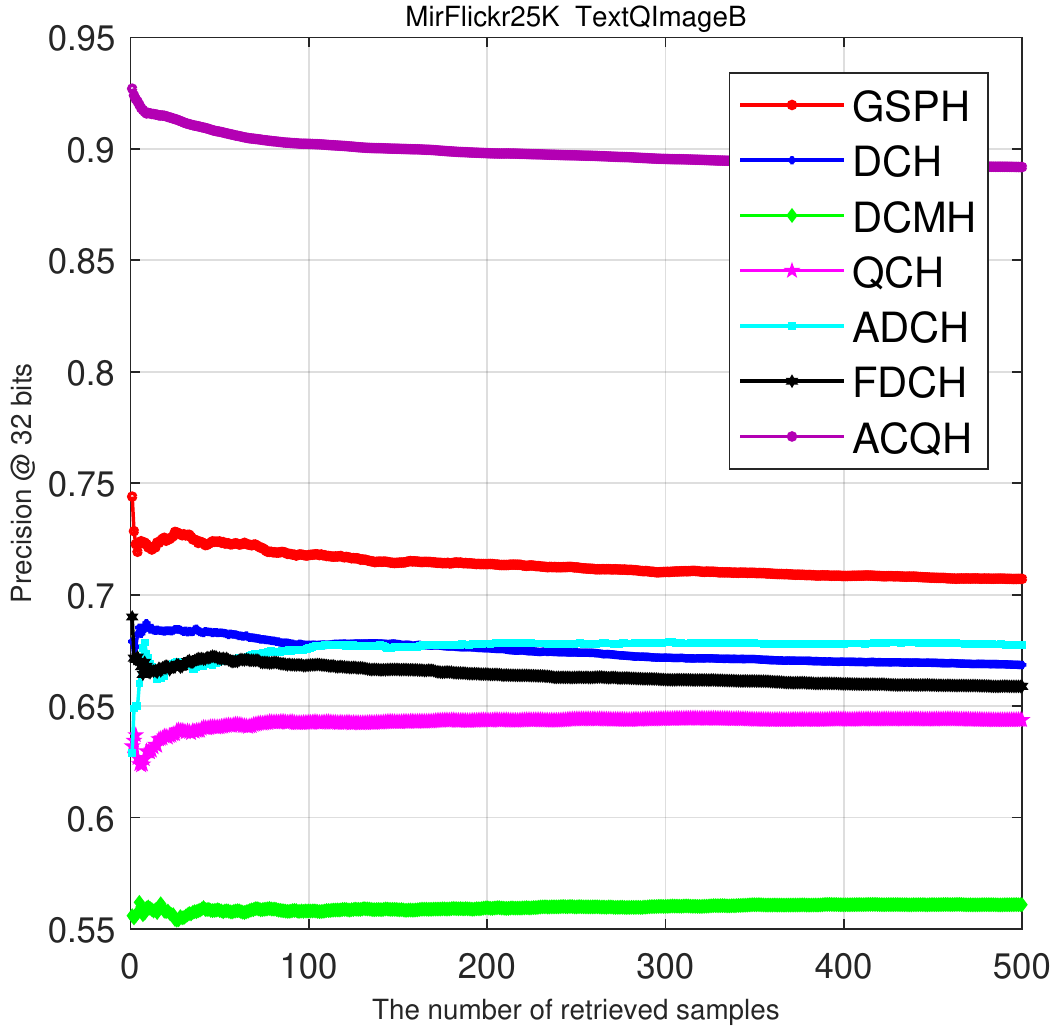}}
  \captionsetup{justification=centering}
\caption{TopN-precision Curves @ $32$ bits on MirFlickr25K.}
\label{figure12}
\end{figure}

\begin{figure} [t]
\centering
\subfigure{
  \includegraphics[width=.2\textwidth]{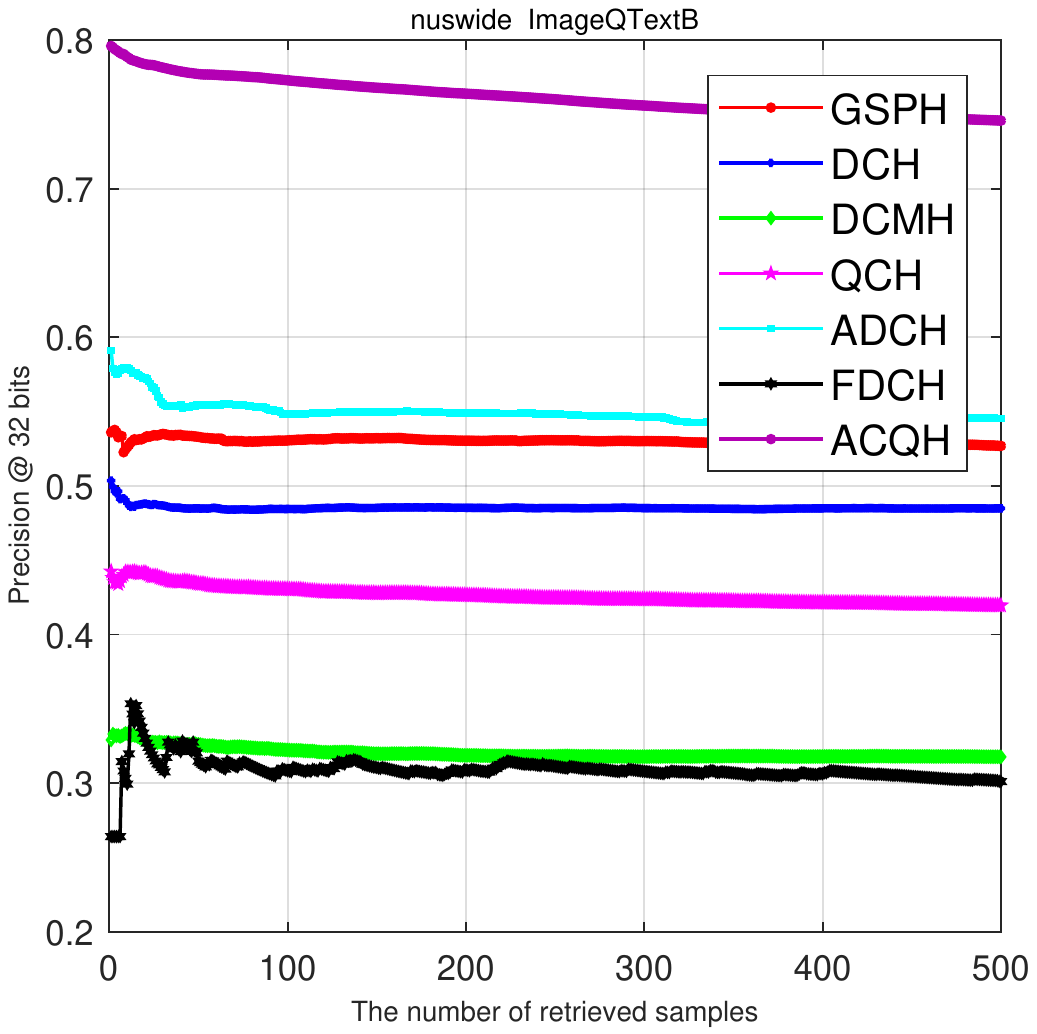}}
\hfill
\centering
\subfigure{
  \includegraphics[width=.2\textwidth]{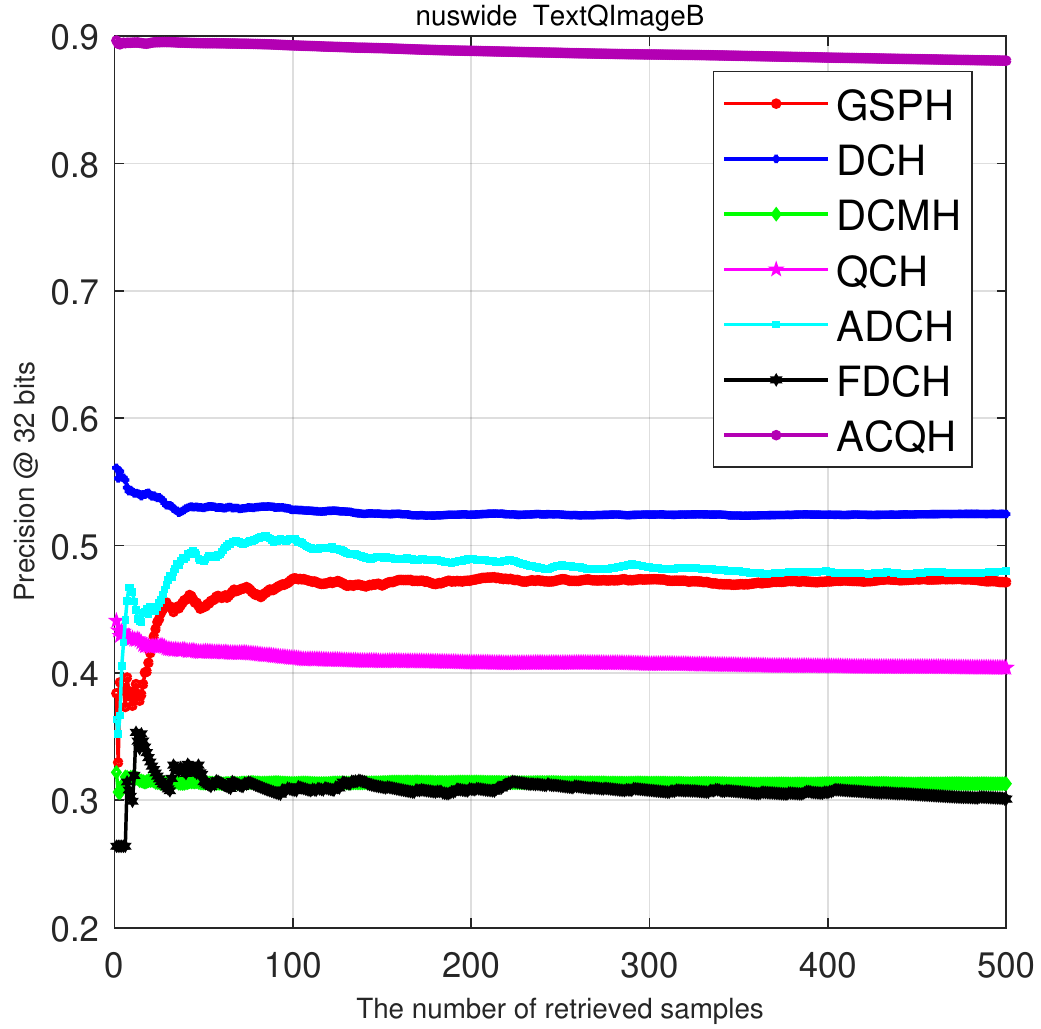}}
  \captionsetup{justification=centering}
\caption{TopN-precision Curves @ $32$ bits on Nuswide.}
\label{figure11}
\end{figure}

\begin{figure} [t]
\centering
\subfigure{
  \includegraphics[width=.2\textwidth]{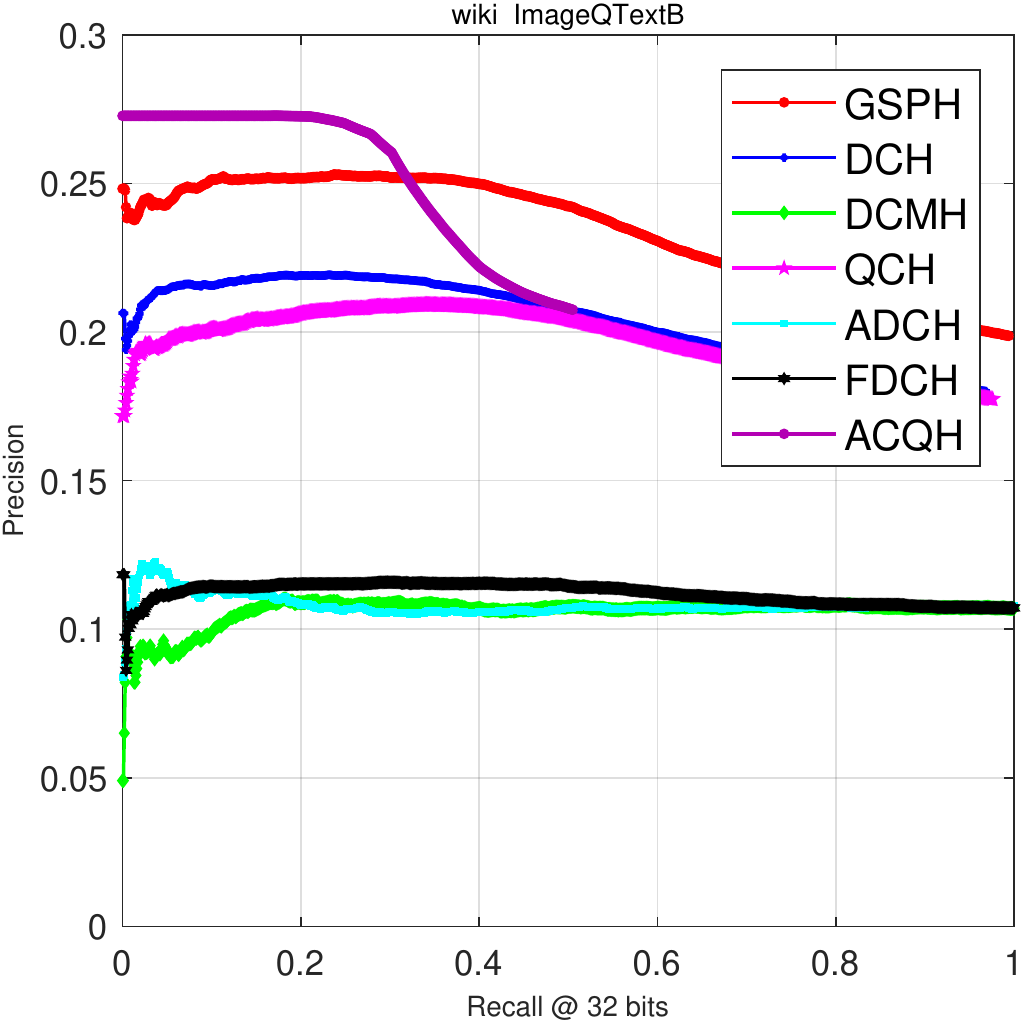}}
\hfill
\centering
\subfigure{
  \includegraphics[width=.2\textwidth]{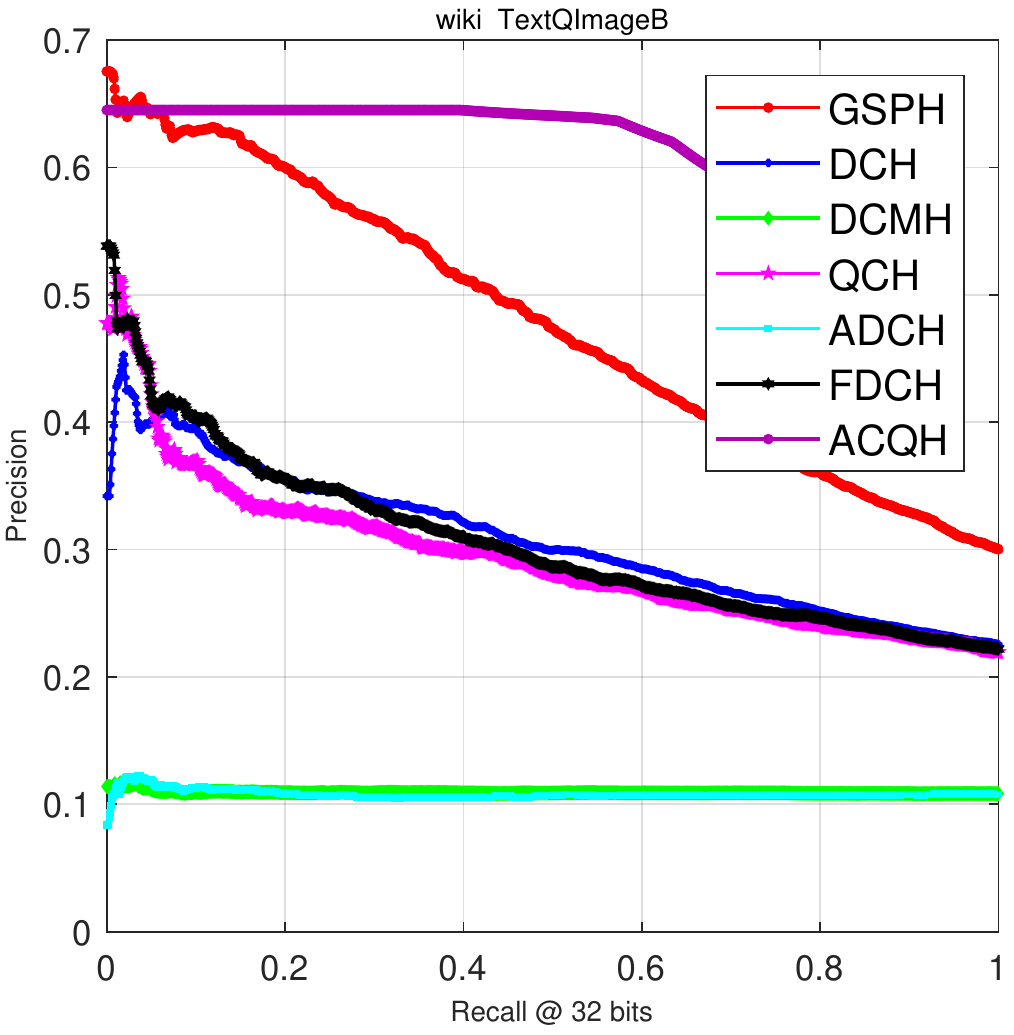}}
\captionsetup{justification=centering}
\caption{Precision-Recall Curves @ $32$ bits on Wiki.}
\label{figure10}
\end{figure}

\begin{figure} [t]
\centering
\subfigure{
  \includegraphics[width=.2\textwidth]{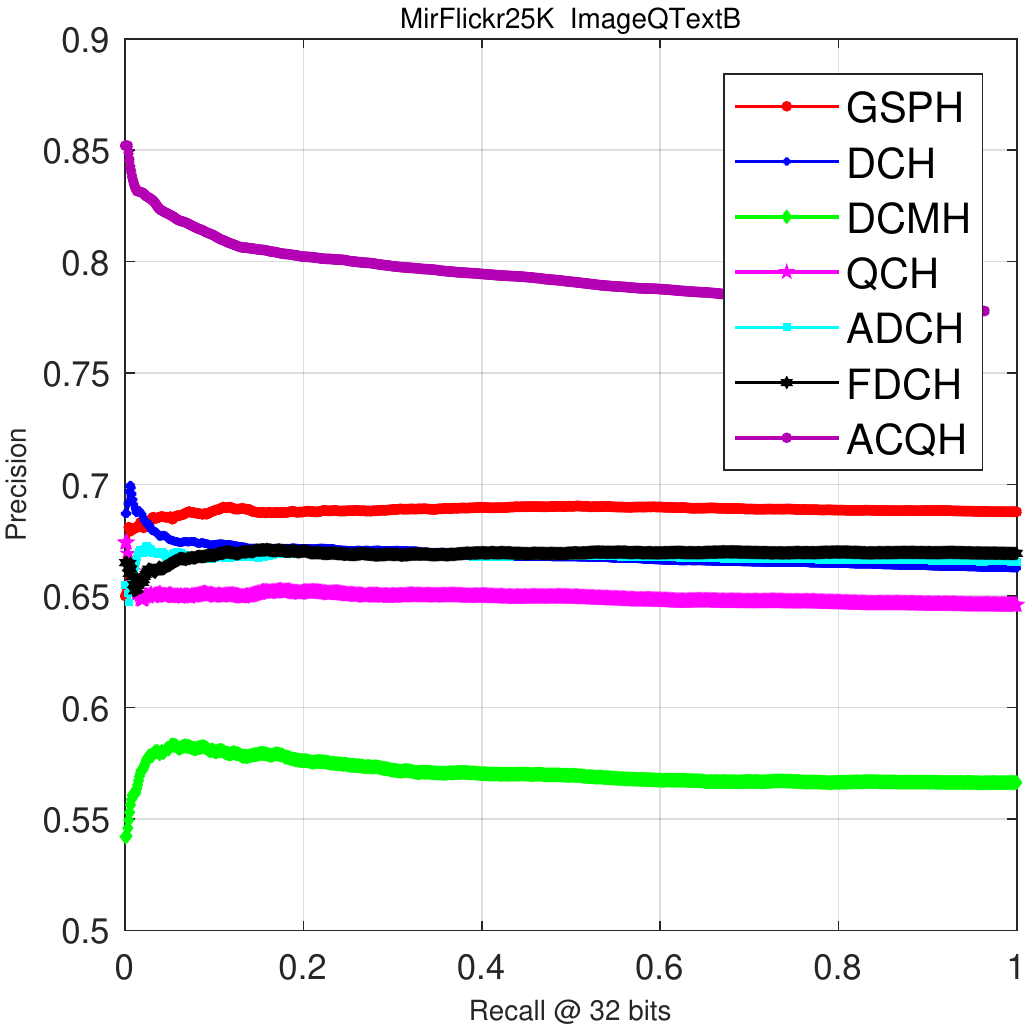}}
\hfill
\centering
\subfigure{
  \includegraphics[width=.2\textwidth]{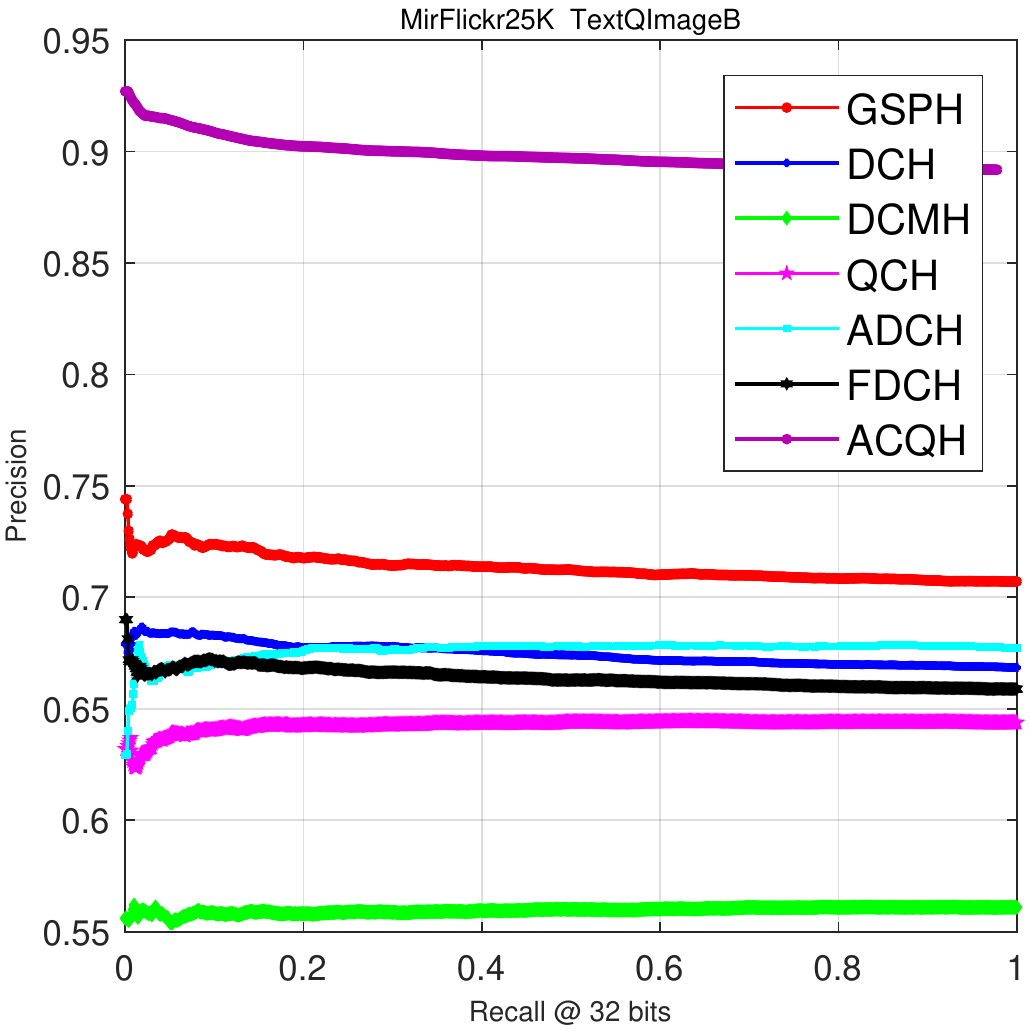}}
\captionsetup{justification=centering}
\caption{Precision-Recall Curves @ $32$ bits on MirFlickr25K.}
\label{figure9}
\end{figure}

\begin{figure}[t]
\centering
\subfigure{
  \includegraphics[width=.2\textwidth]{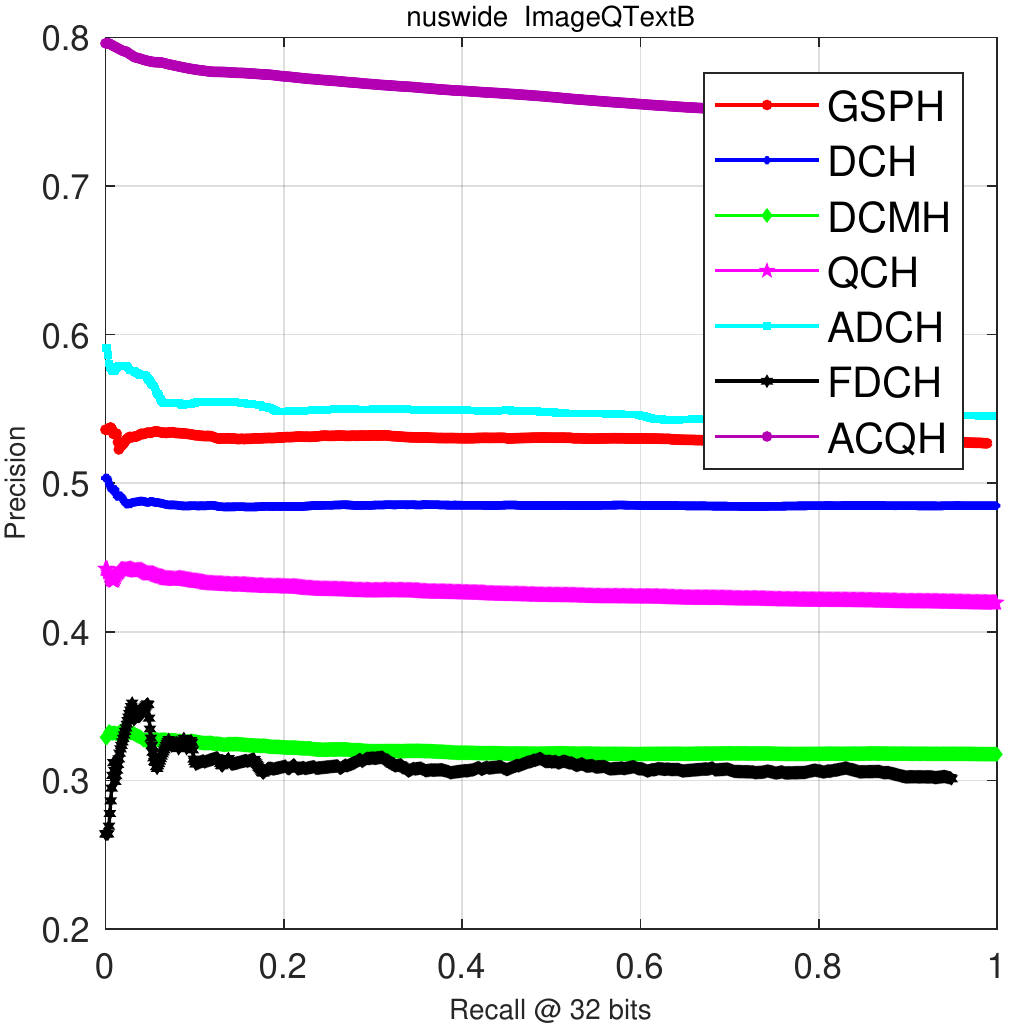}}
\hfill
\centering
\subfigure{
  \includegraphics[width=.2\textwidth]{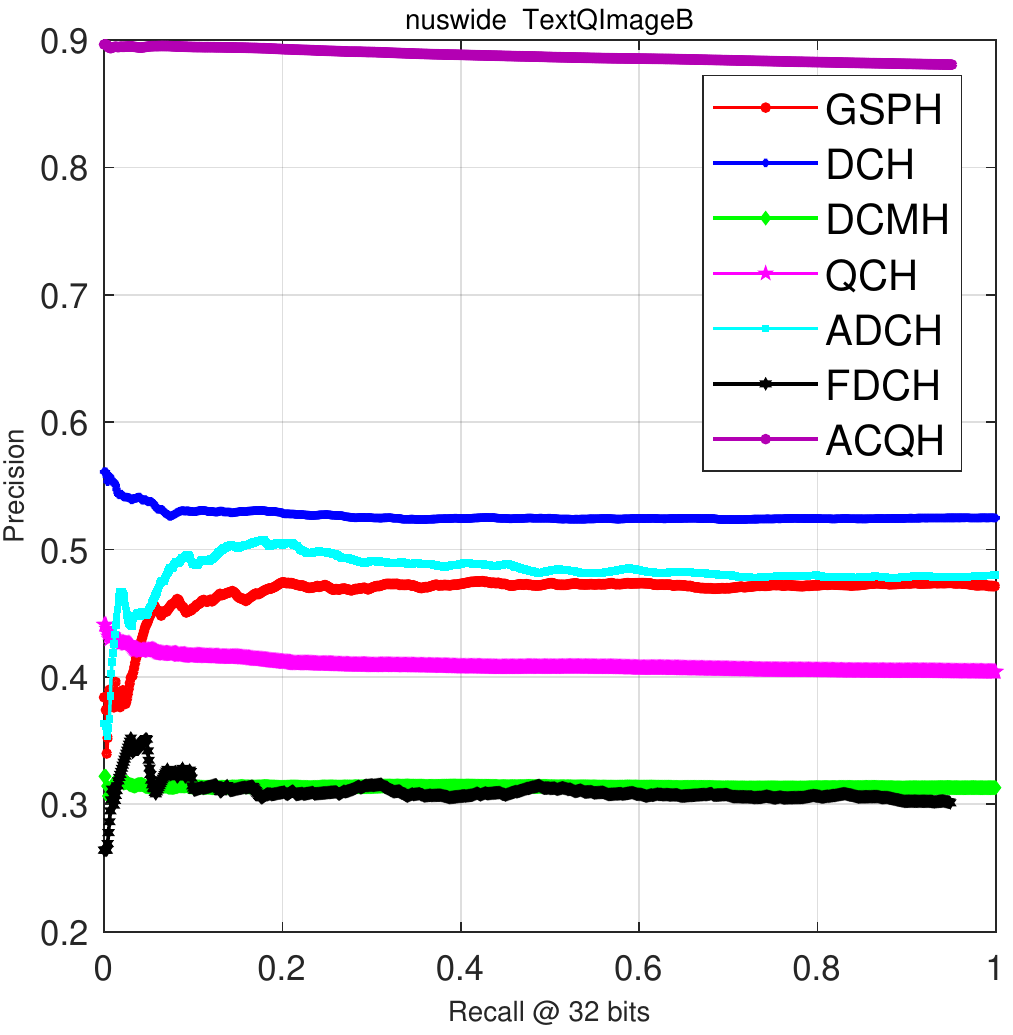}}
  \captionsetup{justification=centering}
\caption{Precision-Recall Curves @ $32$ bits on Nuswide.}
\label{figure8}
\end{figure}

In addition to MAP results, the topN-precision curves at $32$ bits code on all three datasets (i.e. Wiki, MIRFlickr25K, and NUS-WIDE) are present in Figs. \ref{figure13}, \ref{figure12}, and \ref{figure11}, respectively. The topN-precision results show consistently with MAP evaluations, achieving preferable  performance at the front of the search list on all cross-modal retrieval tasks. Besides, the precision-recall curves at $32$ bits are shown in Fig. \ref{figure10}, \ref{figure9}, and \ref{figure8}, respectively. From these figures, we can discover that our ACQH remarkably outperforms the all six baseline CMH methods, further demonstrating the superiority of ACQH.

\renewcommand{\arraystretch}{1.0}
\begin{table*}[htb]
  \centering
  \footnotesize
  \setlength{\belowcaptionskip}{10pt}
  \captionsetup{justification=centering}
  \caption{Training time of different hashing methods on MIRFlickr25K and NUS-WIDE.}
  \label {Table.4}
  \resizebox{0.9\textwidth}{!}{
    \begin{tabular}{|c|c|c|c|c|c|c|c|c|c|c|c|}
    \hline
    \multirow{2}{*}{Method} &
    \multicolumn{5}{c|}{MIRFlickr25K} &\multicolumn{5}{c|}{NUS-WIDE}\cr\cline{2-11}
    &8 bits&16 bits&32 bits&64 bits&128 bits&8 bits&16 bits&32 bits&64 bits&128 bits\cr
    \hline
    GSPH &114.7069	&252.494	&475.0317	&924.0803	&1809.085 &139.8224	&312.6484	&538.0747	&988.1892 &1934.8084
\cr\cline{1-11}
   DCH &1.2047 &3.0871	&6.1125	&28.4328	&130.2607		&5.5003	&7.7615	&9.966	&26.2378	&101.3781
\cr\cline{1-11}
    DCMH &16.4778	&45.5013	&156.2621	&794.0413	&2615.3125	&23.1004	&47.5621	&160.6834	&875.2993	&2709.0421
\cr\cline{1-11}
    QCH &2092.9111	&2151.7798	&2254.7192	&2159.9231	&3431.3842	&1076.2956	&1137.115	&1324.2135	&1463.2668	&1816.6906
\cr\cline{1-11}
    ADCH &4.1281	&8.3598	&17.9763	&40.196	&108.7111		&6.7611	&22.5178	&18.6253	&42.9331	&111.2696
\cr\cline{1-11}
    FDCH &0.3719	&0.4183	&0.4688	&0.5487	&0.8218		&15.5289	&3.7857	&2.5995	&3.6588	&5.688
\cr\cline{1-11}
    {\bf ACQH} &25.6406	&25.8478	&26.2903	&25.3813	&26.2767		&278.7943	&275.0239	&279.1021	&277.1392	&283.8043
\cr
    \hline
    \end{tabular}}
\end{table*}

$\textbf{Training time.}$ In order to test the efficiency of our model ACQH, we carry out experiments on two large benchmark datasets MIRFlickr25K and NUSWIDE to check the time complexity with all the compared CMH methods for training phase. The comparison result of training time complexity is reported in Table. \ref{Table.4}. From this table, we can see that some CMH methods consume a lot time for directly utilizing the $N \times N$ semantic similarity matrix $S$ (i.e. GSPH, QCH, DCMH). Besides, compared with DCH, ADCH, and FDCH, our ACQH spends a little longer time than these methods with shorter binary codes but takes shorter time to DCH, ADCH and longer time to FDCH with longer binary codes. On the other hand, the acceptable training time taking by our approach ACQH is not nearly change with various code length from $8$ bits to $128$ bits, further demonstrating the superiority of ACQH in training aspect.

\begin {figure} [ht]
\centering
\includegraphics[width=.2\textwidth]{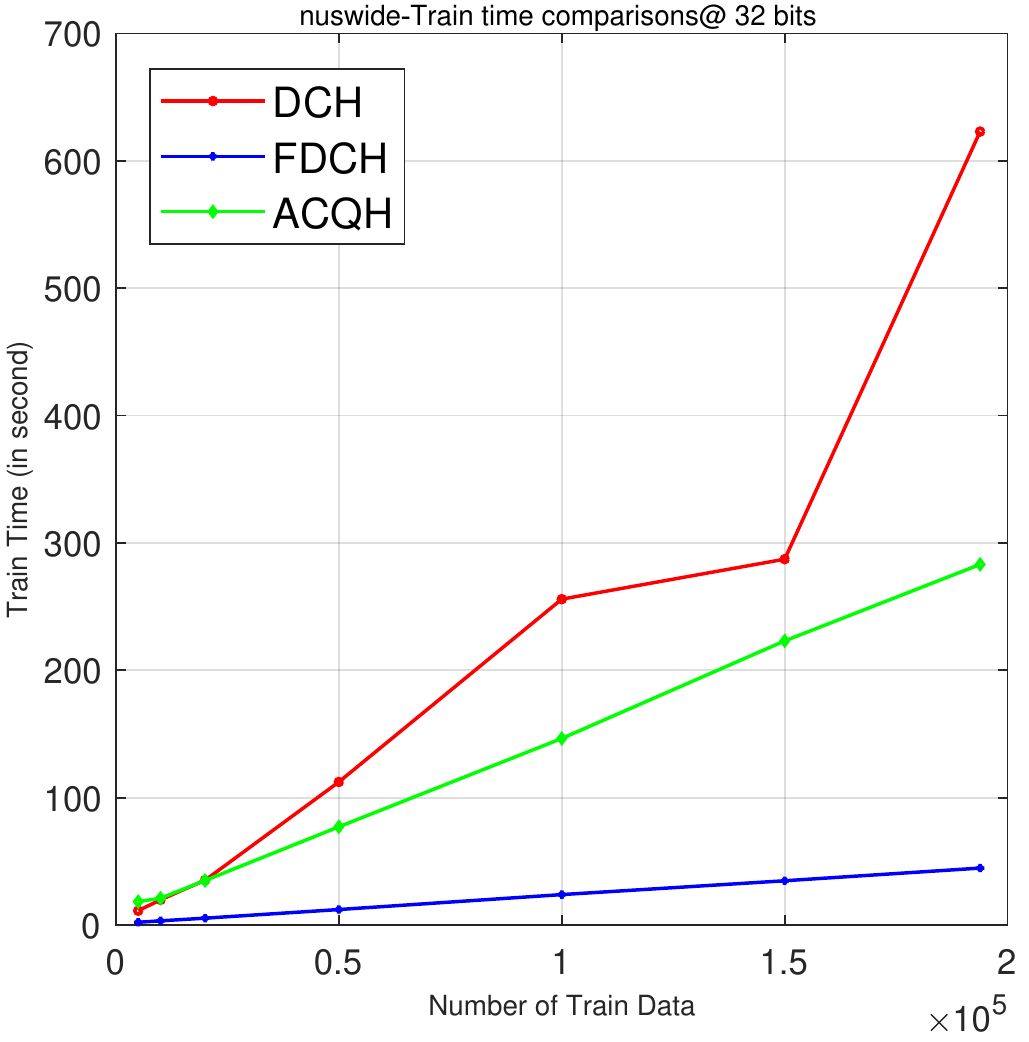}
\caption{Training time (s) on different sizes of subsets in NUS-WIDE at $32$ bits.}
\label{figure4}
\end {figure}

Next, we report the scalability of ACQH at $32$ bits codes in Figs. \ref{figure4} with different training set size from NUS-WIDE dataset, compared with DCH and FDCH (Because ADCH takes more training time than DCH, we ignore ADCH in training time comparison for simplicity). From Figs. \ref{figure4}, we can discover ACQH has better scalability than DCH and has worse scalability than FDCH. However, our approach ACQH shows a linear relation to the training set size. Referring to above conclusion, we can get that ACQH has scalability to large size datasets and long code lengths.

\begin{table*}[htb]
  \centering
  \footnotesize
  \setlength{\belowcaptionskip}{10pt}
  \captionsetup{justification=centering}
  \caption{Testing time (in seconds) of different hashing methods on MIRFlickr25K and NUS-WIDE.}
  \label {Table.5}
  \resizebox{0.9\textwidth}{!}{
    \begin{tabular}{|c|c|c|c|c|c|c|c|c|c|c|c|c|}
    \hline
    \multirow{2}{*}{Tasks}& \multirow{2}{*}{Method} &
    \multicolumn{5}{c|}{MIRFlickr25K} &\multicolumn{5}{c|}{NUS-WIDE}\cr\cline{3-12}
    &&8 bits&16 bits&32 bits&64 bits&128 bits&8 bits&16 bits&32 bits&64 bits&128 bits\cr
    \hline
    \multirow{7}{*}{I$\rightarrow$T}
    &GSPH &0.1996	&0.1933	&0.3131	&0.7145	&1.2874	&5.9816	&7.0509	&9.3775	&15.1869	&27.3455
\cr\cline{2-12}
    &DCH &0.1937	&0.1932	&0.3084	&0.7376	&1.4192	&5.9647	&6.9962	&9.3305	&15.1385	&27.2262
\cr\cline{2-12}
    &DCMH &0.2054	&0.1907	&0.3353	&0.7549	&1.4079	&5.9506	&7.1169	&9.5563	&15.1914	&28.9276
\cr\cline{2-12}
    &QCH &0.1941	&0.186	&0.3269	&0.7384	&1.4318	&5.9592	&7.0046	&9.6071	&15.3215	&29.959
\cr\cline{2-12}
    &ADCH &0.1994	&0.19	&0.3296	&0.7111	&1.3396	&5.9956	&7.0136	&9.7197	&15.1673	&27.856
\cr\cline{2-12}
    &FDCH &0.189	&0.1904	&0.3265	&0.7444	&1.3474	&5.9317	&6.9889	&9.6305	&15.4117	&27.7093
\cr\cline{2-12}
    &{\bf ACQH} &0.1742	&0.176	&0.1778	&0.1793	&0.1777		&3.4695	&3.5356	&3.5013	&3.4951	&3.5988
\cr
  \hline
    \multirow{7}{*}{T$\rightarrow$I}
    &GSPH &0.1519	&0.1993	&0.3043	&0.6873	&1.3014	&6.2943	&7.9643	&9.5588	&15.5335	&28.1106
\cr\cline{2-12}
   &DCH &0.1403	&0.1946	&0.3114	&0.7301	&1.4046	&6.3043	&7.2254	&9.5458	&15.6016	&27.8703
\cr\cline{2-12}
    &DCMH &0.1564	&0.2126	&0.3483	&0.7615	&1.4381	&6.236	&7.2739	&9.5599	&15.447	&27.7086
\cr\cline{2-12}
    &QCH &0.1437	&0.1858	&0.3213	&0.7645	&1.3486	&6.1543	&7.1273	&9.6864	&15.302	&27.6207
\cr\cline{2-12}
    &ADCH &0.1553	&0.196	&0.3348	&0.7615	&1.3431	&6.1926	&7.2224	&9.5632	&15.2971	&28.0253
\cr\cline{2-12}
    &FDCH &0.1374	&0.1847	&0.2939	&0.7374	&1.3568	&5.9666	&6.9965	&9.5546	&21.7061	&28.8588
\cr\cline{2-12}
    &{\bf ACQH} &0.1723	&0.1712	&0.1699	&0.1794	&0.1693		&3.5422	&3.5275	&3.4935	&3.4832	&3.5079
\cr
    \hline
    \end{tabular}}
\end{table*}

$\textbf{Testing time.}$ We compare the testing time with all the compared CMH methods in Table. \ref{Table.5}. From this table, we observe that the compared cross-modal hashing methods take nearly identical time for retrieval but our ACQH takes shorter time than these methods. This is because ACQH construct the similarity matrix by the look-up table operation.
\subsection{Empirical Analysis}
\begin {figure} [ht]
\centering
\includegraphics[width=.2\textwidth]{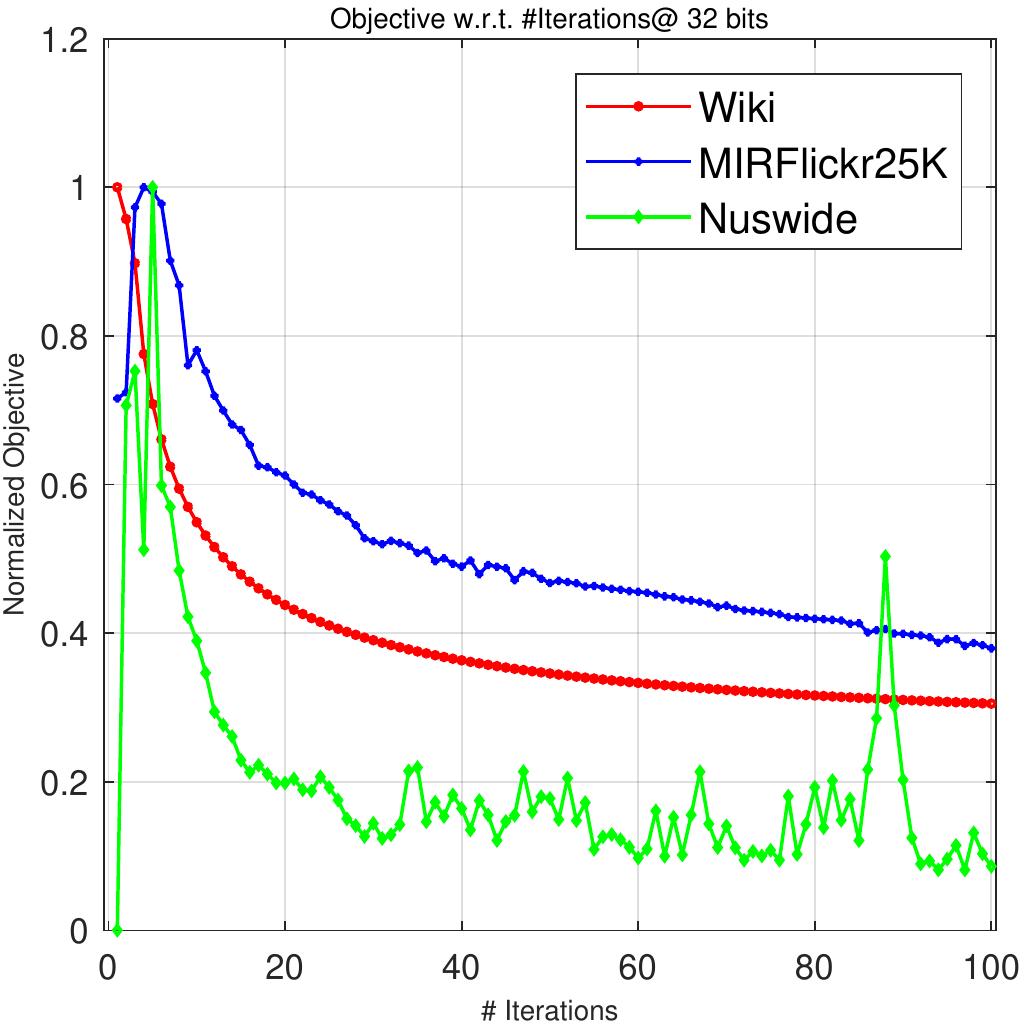}
\caption{Convergence Analysis.}
\label{figure1}
\end {figure}

$\textbf{Convergence.}$ The convergence property of ACQH is validated by empirical analysis. Fig. \ref{figure1} shows the convergence curves on all three datasets at $32$ bits. For convenient observation, the value of the objective is normalized by the number of training data and the maximum in computation. From Fig. \ref{figure1}, ACQH converges with less than $40$ iterations on three datasets, verifying the fast convergence of Algorithm \ref{alg:ACQH}.

\begin {figure} [t]
\centering
\subfigure{
  \includegraphics[width=.2\textwidth]{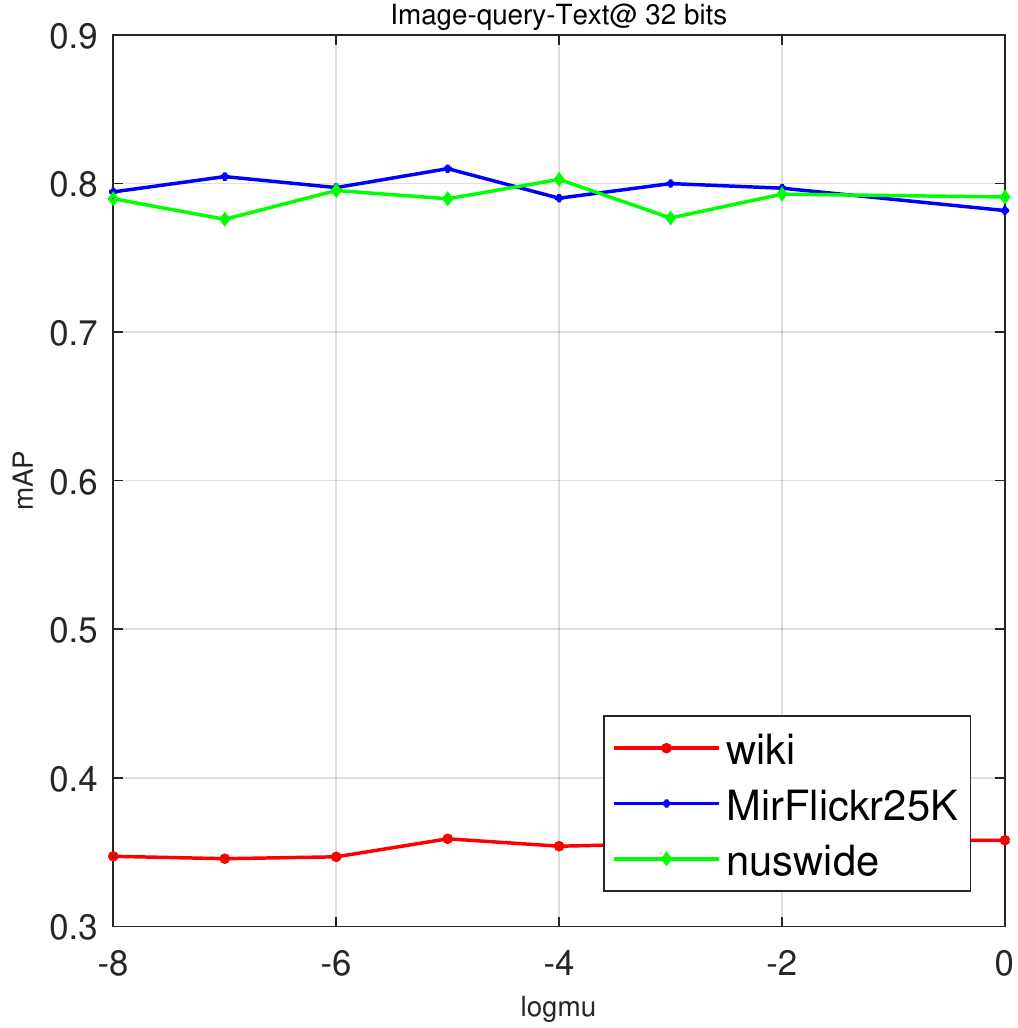}}
\hfill
\centering
\subfigure{
  \includegraphics[width=.2\textwidth]{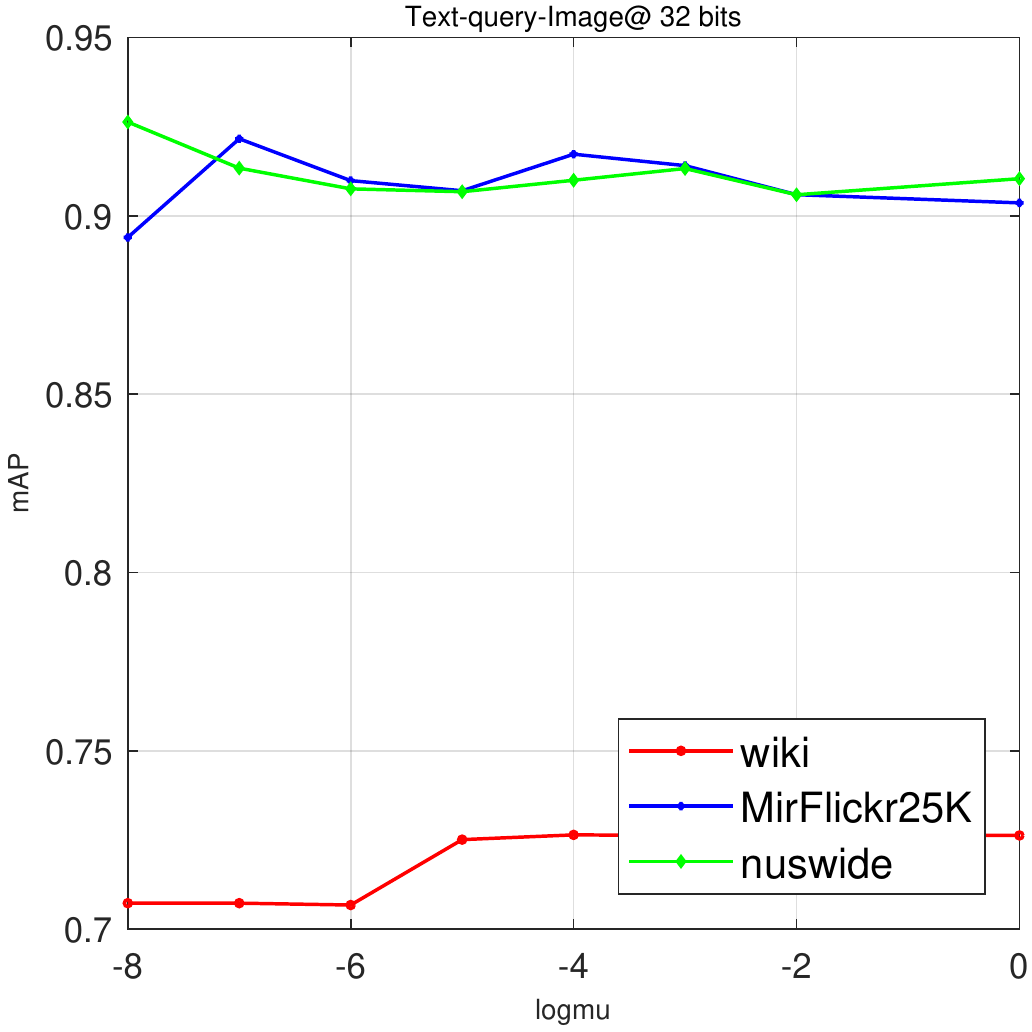}}
\caption{MAP values versus parameter $\mu $.}
\label{figure2}
\end {figure}
\begin {figure} [t]
\centering
\subfigure{
  \includegraphics[width=.2\textwidth]{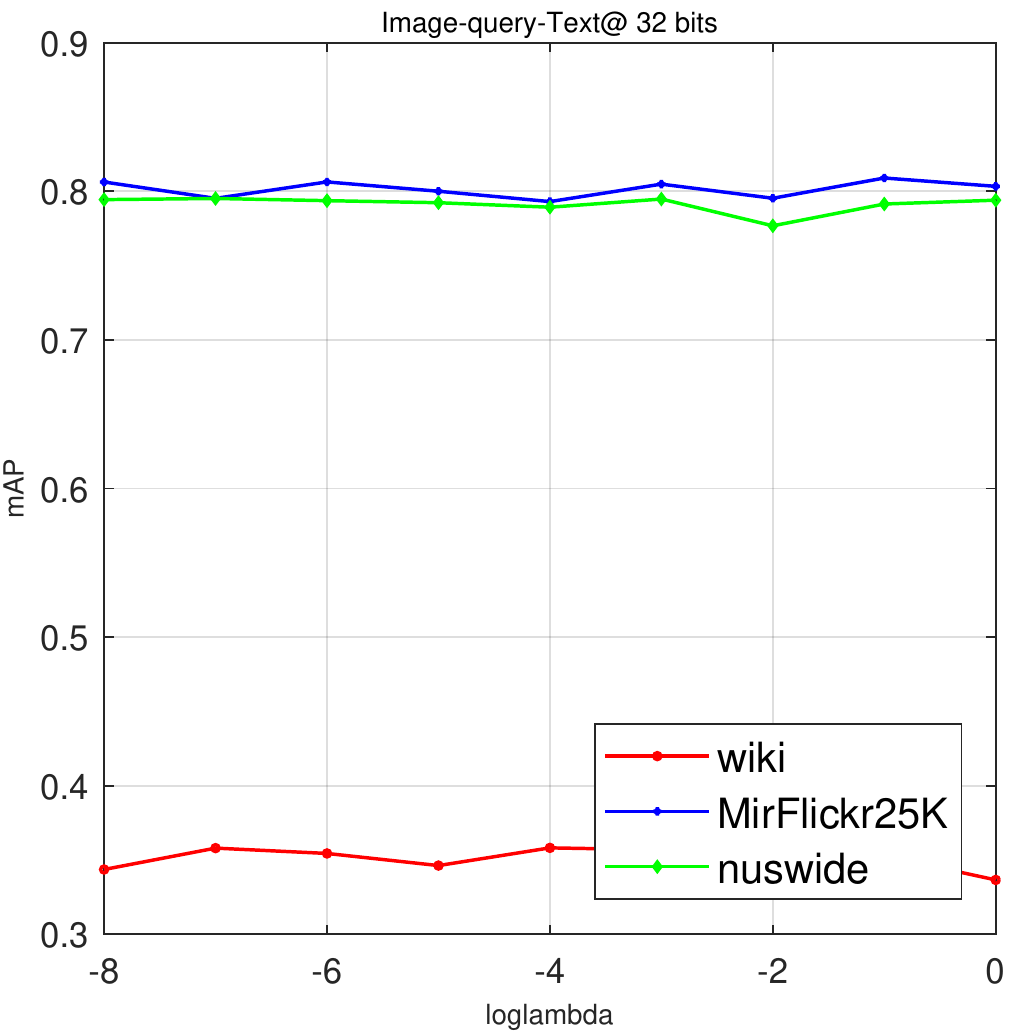}}
\hfill
\centering
\subfigure{
  \includegraphics[width=.2\textwidth]{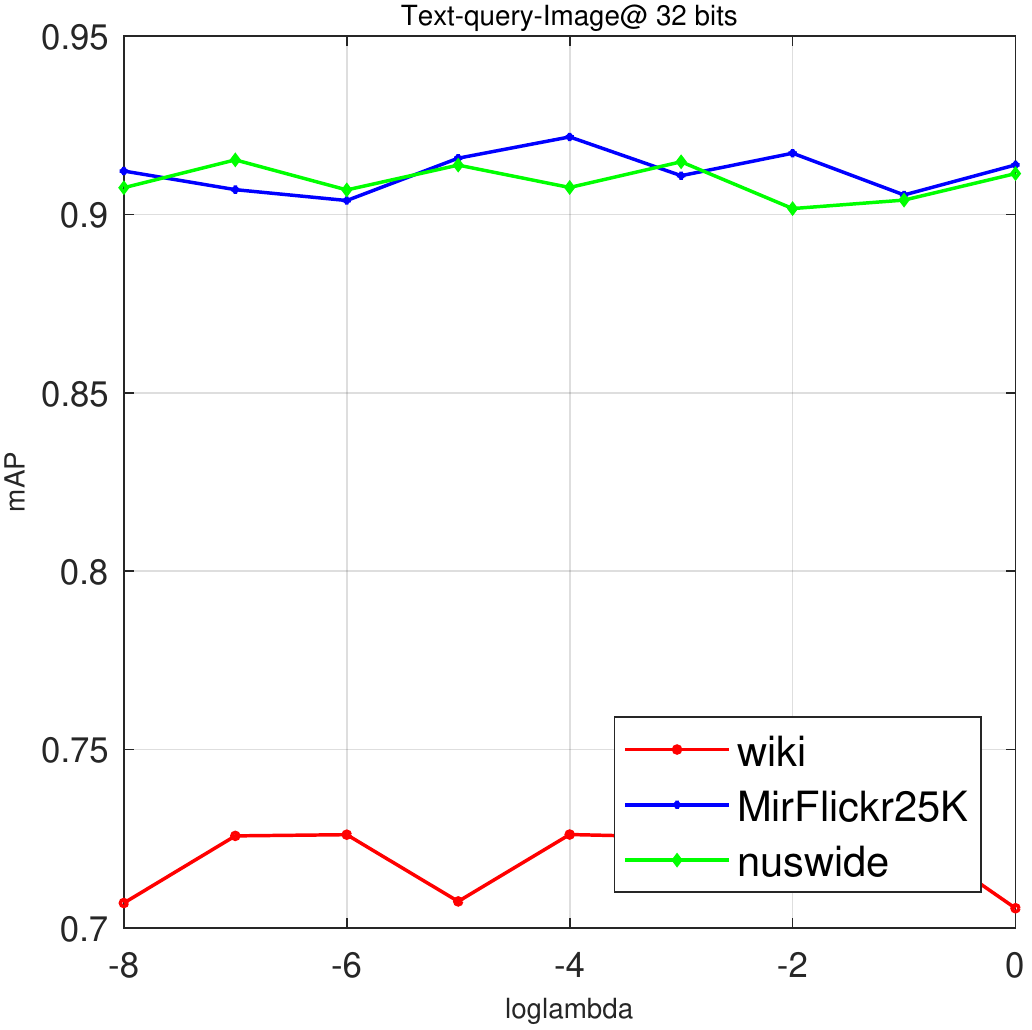}}
\caption{MAP values versus parameter $\lambda $.}
\label{figure3}
\end {figure}
\begin {figure} [t]
\centering
\subfigure{
  \includegraphics[width=.2\textwidth]{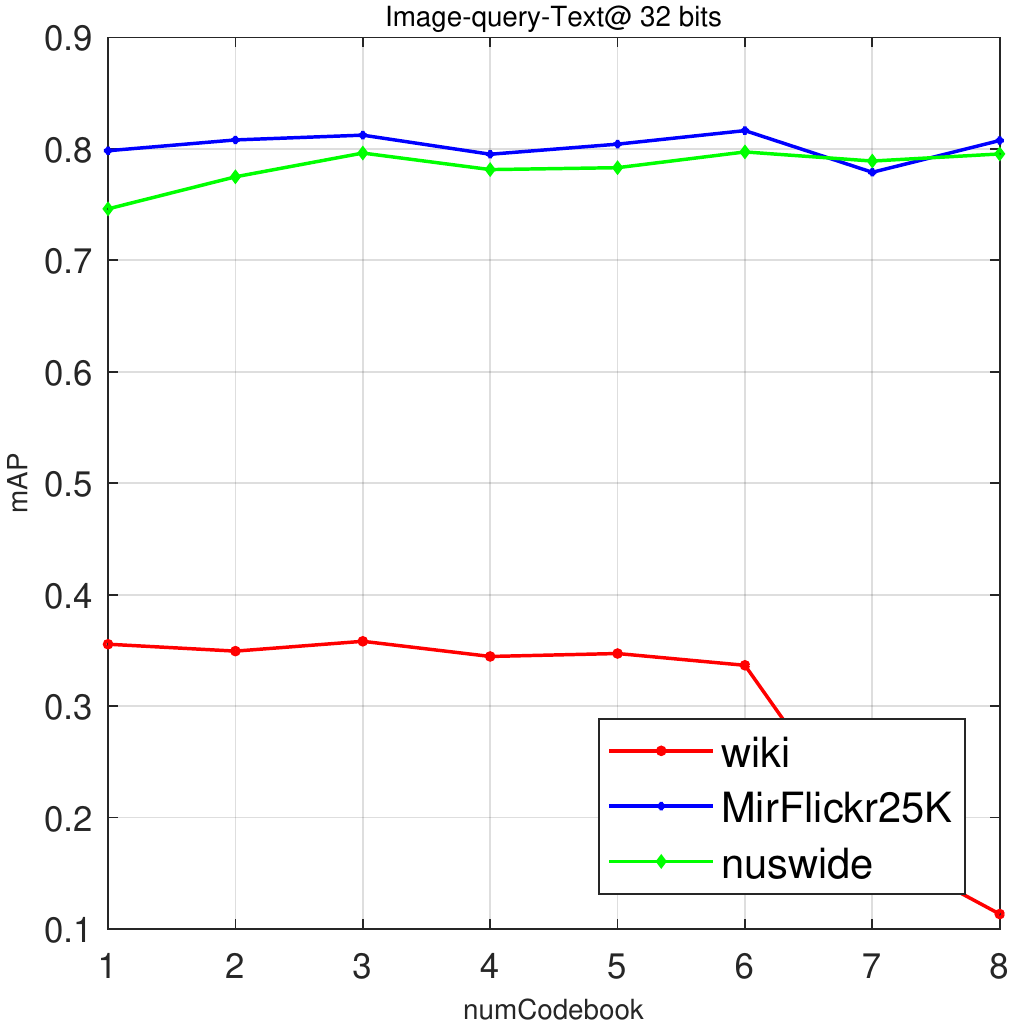}}
\hfill
\centering
\subfigure{
  \includegraphics[width=.2\textwidth]{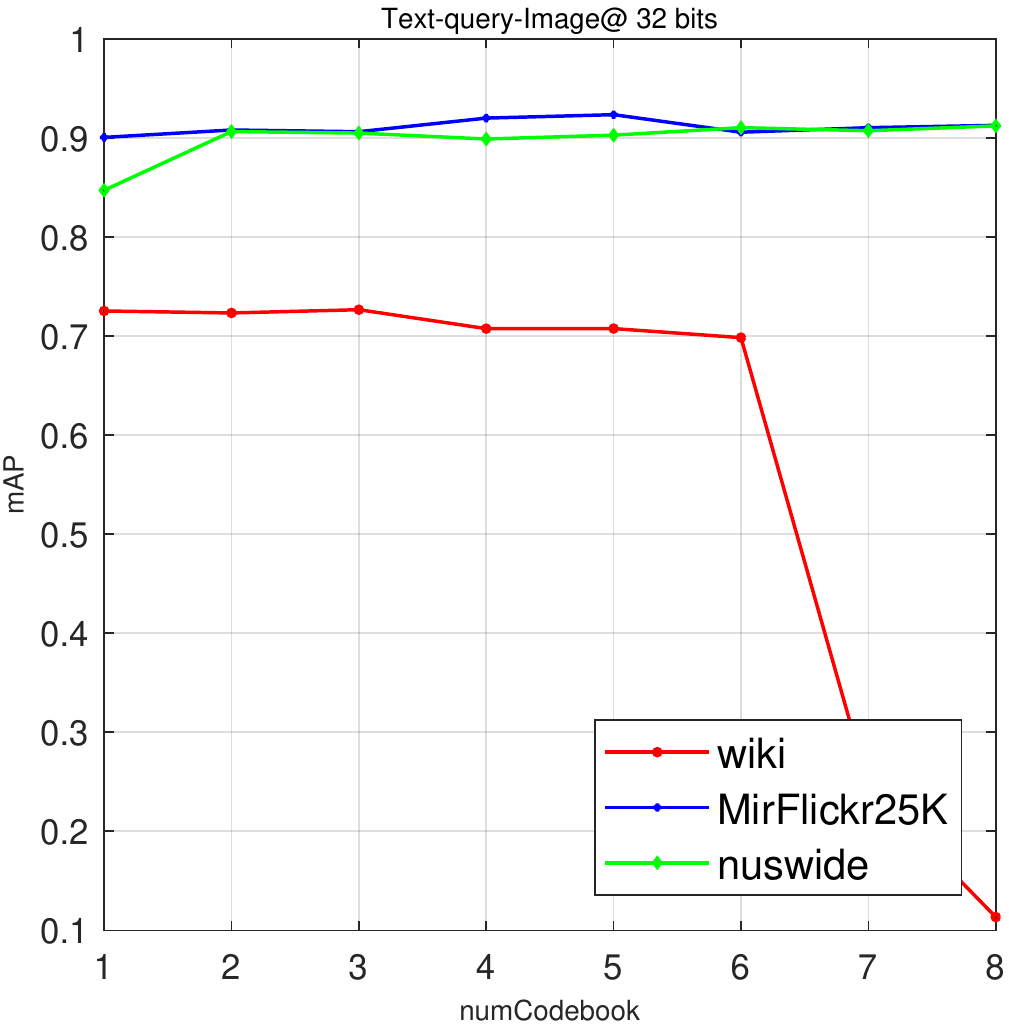}}
\caption{MAP values versus parameter $numCodebook$.}
\label{figure20}
\end {figure}

$\textbf{Parameter Sensitivity Analysis.}$ In this section, We will empirically analyse the parameter sensitivity in different settings with two cross-modal search tasks and the fixed $32$ bits binary codes on all datasets, verifying that our ACQH has superior and stable retrieval results on a large parameter range. We report the MAP results for evaluating the retrieval performance. Our ACQH has three model hyper-parameters, which consists of the regularization hyper-parameter $\mu$, the label regressing error hyper-parameter $\lambda$, the number of sub-dictionaries
hyper-parameter numCodebook. We study them one by one with other hyper-parameters fixed.

The hyper-parameter $\mu$ can avoid the over-fitting about label regressing in ACQH model. From Fig. \ref{figure2}, we can observe that the MAP results of ACQH is not nearly change when $\mu$ increasing, verifying the robust performance of ACQH. We discover ACQH can obtain best performance at $\mu=0.01$ on all datasets.

The hyper-parameter $\lambda$ balances the label regressing error and the pairwise cross-modal asymmetric binary codes learning construction error in ACQH model. From Fig. \ref{figure3}, we can find ACQH can obtain best performance at $\lambda=0.0001$ on all datasets, achieving stable and superior performance on a large parameter range of $\lambda$.

The hyper-parameter numCodebook controls the number of sub-dictionaries in ACQH model. From Fig. \ref{figure20}, we can see that ACQH obtains the best at $numCodebook =4$, having stable and superior performance on a large parameter range of numCodebook.
\section{Conclusion}
In this paper, we have proposed a novel approach for efficient cross-modal similarity search tasks, termed Asymmetric Correlation Quantization Hashing (ACQH) in an asymmetric learning framework. It jointly constructs the mapping matrixes across heterogeneous modalities for projecting query as low-dimensional real-valued vector in latent semantic space and finds the stacked compositional quantization embedding in a coarse-to-fine manner for representing database by a series of learnt real-valued codeword in the codebook with the help of pointwise label information regression, which can lead to significantly improvement of performance and reduction of computation complexity in learning process. Further more, a well-designed discrete iterative optimization framework has been developed to directly learn the unified hash codes  for database points across modalities. Comprehensive experiments on diverse three benchmark datasets have demonstrated the effectiveness and rationality of ACQH, compared with several the state-of-the-art methods. In the future, utilizing non-linear feature mapping such as kernel feature mapping and deep neural network learning is a direction of performance improvement.

\ifCLASSOPTIONcaptionsoff
  \newpage
\fi

\bibliographystyle{IEEEtran}
\bibliography{mybibfile}

\begin{IEEEbiography}[{\includegraphics[width=1in,height=1.25in,
clip,keepaspectratio]{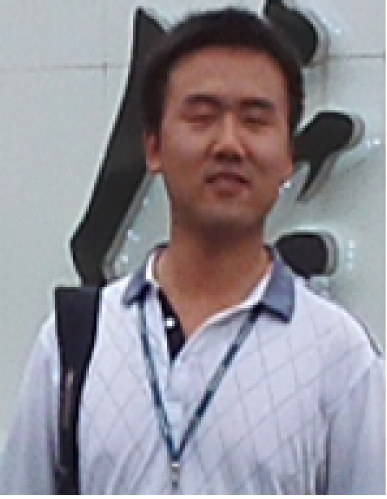}}]{Lu Wang}
received the B.S. degree in electronic information engineering from the Harbin Institute of Technology, Harbin, China, in 2016. He is currently pursuing the Ph.D. degree with the Institute of Image Processing and Pattern Recognition, Shanghai Jiao Tong University, Shanghai, China.

His current research interests include machine learning and information retrieval with respect to learning to hash in Large-scale cross-modal similarity retrieval and visual tracking.
\end{IEEEbiography}

\begin{IEEEbiography}[{\includegraphics[width=1in,height=1.25in,
clip,keepaspectratio]{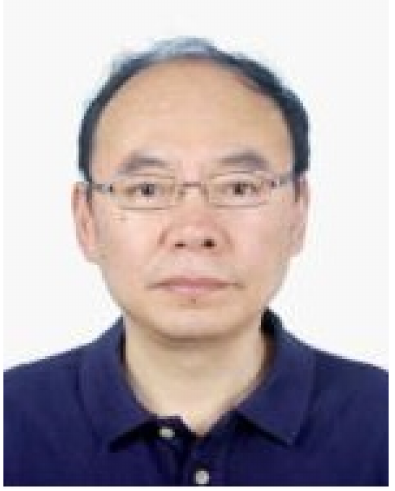}}]{Jie Yang}
received the Ph.D. degree from the Department of Computer Science, Hamburg University, Germany, in 1994.

He is currently a Professor with the Institute of Image Processing and Pattern recognition, Shanghai Jiao Tong University, Shanghai, China. He has been involved in research projects (e.g., National Science Foundation and 863 National High Tech. Plan). He has authored or co-authored one book in Germany, and authored more than 300 journal papers. His current research interests include object detection and recognition, data fusion and data mining, and medical image processing.
\end{IEEEbiography}

\end{document}